\documentclass[10pt,prl,nofootinbib,superscriptaddress,preprintnumbers,balancelastpage,twocolumn]{revtex4-1}
\usepackage{graphicx}
\usepackage{epstopdf}
\usepackage{bm}
\usepackage{bbm}

\usepackage[colorlinks]{hyperref}
\hypersetup{
     colorlinks   = true,
     citecolor    = red,
 linkcolor=blue
}
\usepackage{color}
\usepackage{soul}
\usepackage{amsmath}
\usepackage{amssymb}
\usepackage{mathtools}
\usepackage{xfrac}
\usepackage{ytableau}
\usepackage{rotating}
\usepackage{enumitem}
\usepackage{scrextend}
\usepackage{slashed}

\newcommand{\be}{\begin{eqnarray}}
\newcommand{\ee}{\end{eqnarray}}
\newcommand{\la}{\lambda}
\newcommand{\lt}{\widetilde{\lambda}}

\newcommand{\CC}{\mathbb{C}}

\newcommand{\bu}{{\bf u}}
\newcommand{\bv}{{\bf v}}
\newcommand{\da}{\dot{a}}

\newcommand{\nt}{\widetilde{n}}
\newcommand{\db}{\dot{b}}

\newcommand{\pt}{\widetilde{\partial}}

\usepackage{ulem}
\usepackage{braket}
\usepackage{tcolorbox}
\usepackage[framemethod=tikz]{mdframed}
\usepackage{tablefootnote} 
\makeatletter 
\AfterEndEnvironment{mdframed}{%
 \tfn@tablefootnoteprintout% 
 \gdef\tfn@fnt{0}% 
}
\makeatother 
\newcommand{\bigslant}[2]{{\left.\raisebox{.2em}{$#1$}\middle/\raisebox{-.2em}{$#2$}\right.}}
\def\pd{\partial}
\def\pdt{\widetilde{\partial}}
\def\et{\eta}
\def\etl{\widetilde{\eta}}
\def\scH{\mathcal{H}}
\def\scO{\mathcal{O}}
\def\wtl{\widetilde{l}}
\def\xit{\widetilde{\xi}}
\def\bpi{\boldsymbol{\pi}}
\newcommand*\oline[1]{%
   \vbox{%
     \hrule height 0.5pt%                  % Line above with certain width
     \kern0.5ex%                          % Distance between line and content
     \hbox{%
       \kern-0.0em%                        % Distance between content and left side of box, negative values for lines shorter than content
       \ifmmode#1\else\ensuremath{#1}\fi%  % The content, typeset in dependence of mode
       \kern-0.0em%                        % Distance between content and left side of box, negative values for lines shorter than content
     }% end of hbox
   }% end of vbox
}

\begin{document}

\graphicspath{{Figures/}}

\title{
Conformal-helicity duality \& the Hilbert space of free CFTs
}
\preprint{IPMU19-0015}

\author{Brian Henning}
\affiliation{Department of Physics, Yale University, New Haven, Connecticut 06511, USA}
\affiliation{D\'epartment de Physique Th\'eorique, Universit\'e de Gen\`eve, 24 quai Ernest-Ansermet, 1211 Gen\`eve 4, Switzerland}

\author{Tom Melia}
\affiliation{Kavli Institute for the Physics and Mathematics of the Universe (WPI),The University of Tokyo Institutes for Advanced Study, The University of Tokyo, Kashiwa, Chiba 277-8583, Japan}

\begin{abstract}
  We identify a means to explicitly construct primary operators of free conformal field theories (CFTs) in spacetime dimensions \(d=2\), 3, and 4.
  Working in momentum space with spinors, we find that the \(N\)-distinguishable-particle Hilbert space \(\mathcal{H}_N\) exhibits a \(U(N)\) action in \(d=4\) (\(O(N)\) in \(d=2,3\)) which dually describes the decomposition of \(\mathcal{H}_N\) into irreducible representations of the conformal group. This \(U(N)\) is a natural \(N\)-particle generalization of the single-particle \(U(1)\) little group. The spectrum of primary operators is identified with the harmonics of  \(N\)-particle phase space which, specifically, is shown to be the Stiefel manifold \(V_2(\mathbb{C}^N) = U(N)/U(N-2)\) (respectively, \(V_2(\mathbb{R}^N)\), \(V_1(\mathbb{R}^N)\) in \(d=3,2\)). Lorentz scalar primaries are harmonics on the Grassmannian \(G_2(\mathbb{C}^N) \subset V_2(\mathbb{C}^N)\). We provide a recipe to construct these harmonic polynomials using standard \(U(N)\) (\(O(N)\)) representation theory. We touch upon applications to effective field theory and numerical methods in quantum field theory.
\end{abstract}

\date{\today}

\maketitle
\section{Introduction and summary}

The study of free systems is important in physics, as many interacting systems of interest are deformations of a free theory. Examples include the Standard Model, the Ising model, or essentially any theory defined by a Lagrangian. Two lines of modern research that attempt to exploit this line of thinking are effective field theory (EFT) and Hamiltonian truncation. EFT parameterizes all possible interactions of particles that enjoy the property of being non-interacting when asymptotically separated. In spirit, it is the field-theoretic version of an \(S\)-matrix program; so-called positivity~\cite{Froissart,Gribov:1961ex,Adams:2006sv} and related~\cite{Paulos:2017fhb,Guerrieri:2018uew} results are encouraging hints that it might also admit a bootstrap formulation. In another direction, the decades old idea of Hamiltonian truncation is the field theoretic version of familiar techniques from QM: take a deformation \(\delta H\) of a known, \textit{e.g.} free, Hamiltonian \(H_0\), compute the matrix elements \(\braket{i|\delta H|j}\) with \(\ket{i}\) the Hibert space states of \(H_0\), and diagonalize the matrix. In principle, this non-perturbative approach approximates the spectrum with increasing accuracy as one includes more states. This idea has recently been revived in the high-energy community~\cite{Hogervorst:2014rta,Katz:2016hxp} with encouraging results, \textit{e.g.}~\cite{Hogervorst:2014rta, Katz:2016hxp, Katz:2013qua, Katz:2014uoa, Rychkov:2014eea, Anand:2017yij, Delacretaz:2018xbn}; a central idea in these works is the organization of the states of \(H_0\) according to conformal symmetry.

In both the above examples, crucial ingredients are the operators of the non-interacting theory. In EFT, one focuses on the so-called operator basis which consists of the set of independent operators one can add to the Lagrangian; equivalently, one can think of this as the set of Lorentz preserving deformations of the free theory. In a series of papers, enumeration of the operator basis was solved and steps towards systematic construction taken~\cite{Henning:2015daa, Henning:2015alf, Henning:2017fpj} (see also~\cite{Lehman:2015via,Lehman:2015coa}). While these results follow from Poincar\'e symmetry, it was shown that the conformal group can technically aid in addressing these questions. In Hamiltonian truncation, one needs the basis states, \textit{i.e.} one needs to explicitly construct the primary operators of the unperturbed theory. This is a technically difficult problem, even for free theories. While no systematic construction exists to date, subsets of operators have been constructed and utilized for various problems, \textit{e.g.}~\cite{Craigie:1983fb, Braun:2003rp, Giombi:2016hkj, deMelloKoch:2017caf}.

In this letter we take a significant step towards solving the operator construction problem for free, conformal theories in low spacetime dimensions. Similar to~\cite{Henning:2017fpj,Katz:2016hxp}, we work in momentum space, except here we choose to make use of spinor-helicity variables. The physical idea here is that spinors capture \textit{both} momentum \textit{and} polarization information; therefore---since the spectrum of a free theory is entirely kinematic---one only needs spinors, \textit{i.e.} there is no need to introduce polarization tensors in addition to momentum vectors. Moreover, as we will see, spinor-helicity variables reveal geometric structures in the phase-space which govern the operator content.

In the rest of this section we will explain the basic idea and state our central result, focusing on the case with spacetime dimension $d=4$ (the analogous results for $d=2,3$ are sketched in the supplementary material);  subsequent sections formalize these statements, provide examples, 
and discuss future directions. A companion paper~\cite{companion} elaborates further on scalar primary operators, and construction of the EFT operator basis.

\paragraph{Phase space and the Stiefel manifold} The basic idea is simple: due to the Fock space structure of the Hilbert space, in free theories operators simply consist of polynomials in the fields and their derivatives. Moreover, when the theory is conformal\footnote{Although true in \(d=4\), in general masslessness does not guarantee conformality; see~\cite{siegel} for the classification.} this spans the Hilbert space, due to the operator-state correspondence. In momentum space, the operators translate into polynomials of the momentum and polarization tensors---or, in \(d=4\), simply polynomials in spinor variables \(\la\) and \(\lt\).\footnote{Our conventions can be deduced from the following equations: \(p_{a\da} = p_{\mu}\sigma^{\mu}_{a\da} = \left( \begin{smallmatrix} p_0 + p_3 & p_1 - ip_2 \\ p_1 + ip_2 & p_0 - p_3 \end{smallmatrix} \right)_{a\da} \) with magnitude \(p^2 = g^{\mu\nu}p_{\mu}p_{\nu} =\det p_{a\da} = \frac{1}{2} \epsilon^{ab}\epsilon^{\da\db}p_{a\da}p_{b\db} = \frac{1}{2} p_{a\da}p^{a\da}\).}

Now consider \(N\) distinguishable particles carrying total momentum
\begin{equation}
  P_{a\da} = \sum_{i=1}^N \la^i_{a}\lt_{\da i}^{}.
\end{equation}
In addition to the usual Lorentz group \(SL(2,\mathbb{C})\) action on the spinors, there is generically a \(GL(N,\mathbb{C})\) action rotating the spinors amongst themselves. Importantly, the \(U(N) \subset GL(N,\mathbb{C})\) subgroup---under which \(\la\) transforms in the fundamental (defining) representation and \(\lt\) in the conjugate representation---leaves \(P\) invariant. In this sense the \(U(N)\) action naturally generalizes the various \(U(1)\) little group scalings we can perform on each of the spinors; in particular, the \(N\) little group scalings form the torus, \(U(1)^N \subset U(N)\). 

For a single particle, \(N=1\), the helicity is a representation of the \(U(1)\) little group that leaves the momentum invariant. This generalizes to \(N\) particles: besides carrying some net momentum \(P\), the possible other properties of the system of particles is encoded by the \(U(N)\) in a particular way.

\begin{figure}
\begin{center}
\includegraphics[width=8.5cm]{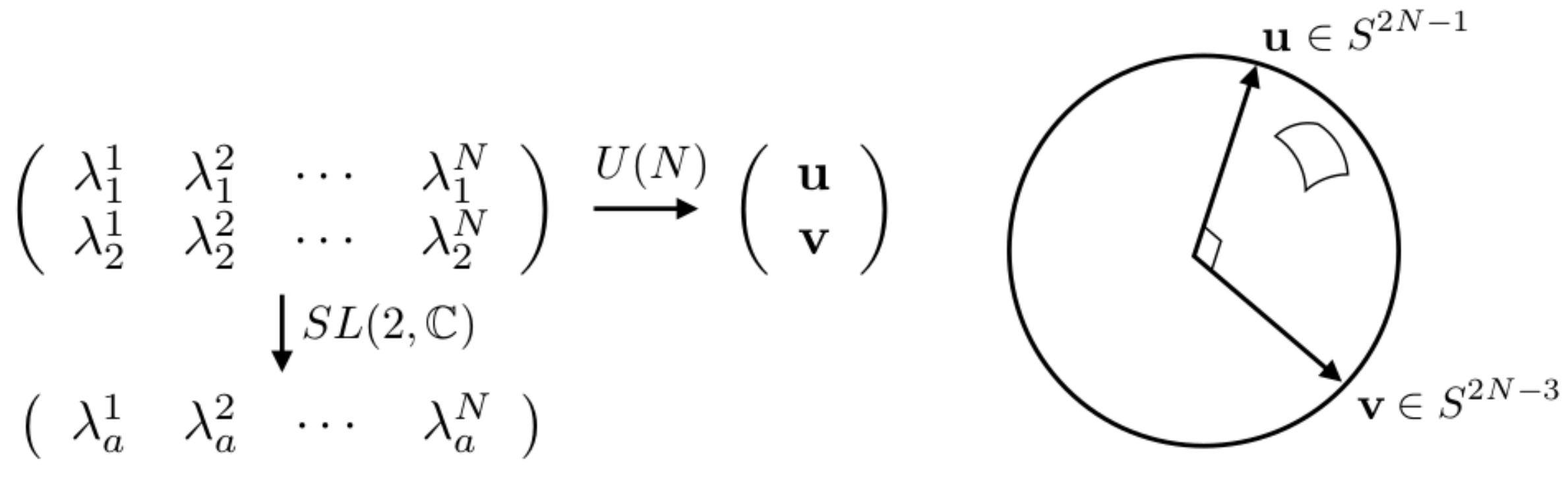}
\end{center}
\caption{{\it Left:} Interpretation of kinematic data with reference to both an $SL(2,\mathbb{C})$ (Lorentz) and a $U(N)$ action. {\it Right:} Geometry of the Stiefel manifold.}
\label{fig:slun}
\end{figure}

More precisely, \(P = \sum \la\lt\) carves out some manifold in \(\mathbb{C}^{2N}\). To elucidate this, write \(\la^i = \begin{pmatrix} u^i \\ v^i \end{pmatrix}\) and go to the center of mass frame \(P_{a\da} = M\delta_{a\da}\), 
\begin{equation}
  \begin{pmatrix} M & 0 \\ 0 & M \end{pmatrix} = \begin{pmatrix} \left|\mathbf{u}\right|^2 & \bv^{\dag} \bu \\ \bu^{\dag} \bv & \left|\bv\right|^2 \end{pmatrix}.
\end{equation}
Evidently, in this frame, the manifold consists of two complex spheres of radius \(\sqrt{M}\) that are tangential to one another (\(\bv^{\dag} \bu =0\)) (see Fig.~\ref{fig:slun}). 
 It is easiest to think of this as a homogeneous space: the two \(U(N)\) vectors \(\bu\) and \(\bv\) take ``vevs'' and ``break'' \(U(N) \to U(N-2)\), so that the manifold is identified with \(U(N)/U(N-2)\). This manifold is known as the (complex) Stiefel manifold of 2-frames, \(V_2(\mathbb{C}^N)\).

Physical observables of the \(N\) particles only have support on the Stiefel manifold. That is, observables are functions \(f(\la_a^i,\lt_{\da i})\) on \(V_2(\mathbb{C}^N) \subset \mathbb{C}^{2N}\). Certain observables---in particular, scattering amplitudes---are Lorentz invariant: \(f(g\la,g^*\lt) = f(\la,\lt)\) for \(g \in SL(2,\mathbb{C})\). In this case, we can mod out by an additional \(U(2)\) corresponding to the little group of the massive momentum \(P\) (and a complex phase), and consider functions on the Grassmann manifold \(G_2(\mathbb{C}^N) = U(N)/U(N-2)\times U(2)\). This is nothing but the ``kinematic Grassmannian''~\cite{ArkaniHamed:2009dn} that proliferates the modern study of four-dimensional scattering amplitudes. Intuitively, the ``more general'' correlation functions live on the ``more general'' Stiefel manifold, while the Lorentz singlet data lives on a submanifold. Mathematically, the Stiefel manifold is a \(U(2)\) fiber bundle over the Grassmannian \(U(2) \to V_2(\CC^N) \to G_2(\CC^N)\); physically, this reflects familiar manipulations like decomposing correlation functions or form factors into Lorentz spin structures times Lorentz invariant functions (see, \textit{e.g.}, \cite{Kravchuk:2016qvl} for a systematic discussion in the context of conformal correlation functions). 

In consideration of observables, the harmonics on the Stiefel/Grassmann manifold provide a natural basis for such functions. It becomes obvious how to ascertain these harmonics upon reviewing the case of spherical harmonics, which we do in the supplemental material (this also provides the analogous story for \(d=2\) dimensions). The upshot is that {\it (i)} these originate from polynomials in the spinors---called {harmonic polynomials}---that are {annihilated} by the generalized Laplacian
\begin{equation}
  K^{a\da} = -\sum_i\pd^a_i\pt^{\da i}
  \label{eq:SC_gen}
\end{equation}
and {\it (ii)} the polynomials furnish certain (finite-dimensional) representations of \(U(N)\).

The physical content of these statements becomes transparent upon recognizing that this generalized Laplacian is the generator of special conformal transformations: \(P\) and \(K\)---together with the Lorentz and dilatation generators\footnote{Respectively (with ``\(\cdot\)'' denoting a sum over Lorentz indices),
  \begin{align}
  M_a^b &=-i \sum_i\Big(\la_a^i\pd^b_i - \frac{1}{2}\delta_a^b \la^i\cdot \pd_i\Big),~~~
  \widetilde{M}_{\da}^{\db} =-i \sum_i\Big(\lt_{\da i}\pt^{\db i} - \frac{1}{2}\delta_{\da}^{\db} \lt_i\cdot \pt^i\Big), \nonumber \\
  D &=-\frac{i}{2} \sum_i\Big(\la^i \cdot \pd_i + \lt_i\cdot \pt^i + 2\Big).
  \end{align}
}
\(M\), \(\widetilde{M}\), and \(D\)---furnish the conformal algebra \(\mathfrak{su}(2,2) \simeq \mathfrak{so}(4,2)\). In particular, the harmonic polynomial condition---annihilation by \(K\)---{implies that the corresponding operator is a primary operator}.

The linking of the harmonic and primary condition implies that \textit{the conformal representation theory is determined by the \(U(N)\) representation theory and vice versa}. This can be shown in numerous ways. One simple way is by showing that the Casimir operators of \(\mathfrak{su}(2,2)\) can be written in terms of the Casimir operators of \(\mathfrak{u}(N)\),\footnote{The representation of the \(\mathfrak{u}(N)\) algebra is given by
  \begin{equation}
    E^i_j =-i\big( \la^i\cdot \pd_j - \lt_j\cdot \pt^i \big),~~~[E^i_j,E^k_l] = -i\big(\delta_j^kE^i_l - \delta^i_lE^k_j\big).
\label{eq:ungen}
  \end{equation}
  As a rank-3 algebra, \(\mathfrak{su}(2,2)\) nominally can have three independent Casimirs. For \(N\ge 3\) all the \(\mathfrak{su}(2,2)\) Casimirs are independent and can be written in terms of the \(\mathfrak{u}(N)\) Casimirs. For \(N=1\), resp. 2, there are only one, resp. two, independent conformal Casimirs; a related fact is that these contain ``short'' representations (the shortening conditions coming from equations of motion, resp. current conservation).
}
and vice-versa. A more constructive method fleshes out the ideas sketched in the previous paragraphs, and will be taken up in the next section. 

As the \(U(N)\) generalizes helicity, we call this link with the conformal group \textit{conformal-helicity duality}. Many of the ideas presented here have appeared in the mathematics literature, where the \(SU(2,2) \times U(N)\) action on the space of polynomials in the spinors is an example of a {reductive dual pair} within the oscillator representation~\cite{howe, howe1985} (see the supplemental material for a brief introduction). We note that the single particle $N=1$ case essentially coincides with the analysis of~\cite{MackTod}. In addition to the general theory developed by Howe~\cite{howe,howe1985}, the papers~\cite{kashiwara1978,gelbart1974} contain relevant results.

Our main result can be summarized as: 
\rule{\linewidth}{0.5pt}
%\begin{mdframed}[leftline=false,rightline=false,frametitle={Main result}]
\begin{addmargin}[1em]{0em}
\noindent{\textbf{Main Result}}

\noindent For free, massless particles of arbitrary helicity in four dimensions, denote by \(\mathcal{H}_N\) the Hilbert space of \(N\) distinguishable particles (\(\mathcal{H}_N = \mathcal{H}_1^{\otimes N}\)). By the operator-state correspondence, we equivalently think of \(\mathcal{H}_N\) as the space of local operators characterizing the particles. %%
\(\scH_N\) furnishes a (reducible) unitary representation of \(SU(2,2) \times U(N)\) whose irreducible decomposition consists of an infinite number of positive-energy irreducble representations (irreps) of \(SU(2,2)\)~\cite{Mack:1975} carrying an irrep of \(U(N)\). The decomposition is such that the \(U(N)\) irrep specifies the \(SU(2,2)\) irrep, and vice-versa.

Using the usual partitions \(L = (L_1,\dots,L_N)\) with \(L_i \in \mathbb{Z}\) and \(L_1 \ge L_2 \ge \cdots \ge L_N\) to label \(U(N)\) irreps, the decomposition is
\begin{equation}\label{eq:decomp_HN}
  \scH_N = \bigoplus_{L \in \Lambda^{(N)}} \mathcal{V}_L,~~~~\mathcal{V}_L = R_{[\Delta,j_1,j_2](L)} \otimes W_{L},
\end{equation}
where \(W_L\) is an irrep of \(U(N)\) labeled by partition \(L\), \(R_{[\Delta,j_1,j_2](L)}\) is a positive-energy irrep of \(SU(2,2)\) with scaling dimension and spin determined from \(L\) (see eq.~\eqref{eq:Dj1j2_from_L} below), and \(\Lambda^{(N)}\) is the set of \(L\)s appearing in the decomposition. 

For \(N \ge 4\) this set is given by
\begin{align}
  \Lambda^{(N\ge 4)} = \Big\{ (l_1,&l_2,0,\dots,0,-\wtl_2,-\wtl_1) ~~\text{such}\nonumber \\
  &\text{that}~~l_i,\wtl_i \in \mathbb{Z}_{\ge 0},~l_1\ge l_2, ~\wtl_1 \ge \wtl_2 \Big\},
  \label{eq:masterweight}
\end{align}
where the negative numbers are a standard notation for \(U(N)\) conjugate representations, \textit{e.g.} sec.~15.5 of~\cite{Fulton}.\footnote{In the physics literature it is common to use an overline to denote conjugate representations: \(\overline{(L_1,\dots,L_N)} = (-L_N,\dots,-L_1)\).} As we will see in the next section, \(l_{1,2}\) (\(\wtl_{1,2}\)) are natural labels for polynomials in \(\la_a^i\) (\(\lt_{\da i}\)) (see Fig.~\ref{fig:glue}).

The corresponding \(SU(2,2)\) quantum numbers are
\begin{align}\label{eq:Dj1j2_from_L}
  \Delta &= \frac{1}{2}\big(l_1 + l_2 + \wtl_1 + \wtl_2\big) + N, \nonumber \\
  j_1 &= \frac{1}{2}\big(l_1 - l_2\big),~~  j_2 = \frac{1}{2}\big(\wtl_1 - \wtl_2\big),
 % j_2 &= \frac{1}{2}\big(\wtl_1 - \wtl_2\big), \nonumber
\end{align}
together with a net helicity quantum number \(h\),\footnote{One can think of this either as the charge of the diagonal \(U(1)\) in \(U(N) \simeq SU(N)\times U(1)\) or as that of \(U(2,2) \simeq SU(2,2) \times U(1)\).}
\begin{equation}
  h = \frac{1}{2}\big(l_1 + l_2 - \wtl_1 - \wtl_2 \big).
\end{equation}

The \(N < 4\) cases are somewhat special; retaining the \(l_{1,2}\) and \(\wtl_{1,2}\) parameters they are
\begin{subequations}\label{eq:exceptional}
\begin{align}
  %\Lambda^{(1)} &= \big\{ (l_1) ~\text{or}~(-\wtl_1)~~:~~l_1,\wtl_1\in \mathbb{Z}_+ \big\}\\
  \Lambda^{(1)} &= \big\{ (l_1) ~\text{and}~(-\wtl_1)\big\},\\
  \Lambda^{(2)} &= \big\{(l_1,l_2),~(l_1,-\wtl_1),~\text{and}~(-\wtl_2,-\wtl_1) \big\}, \\
  \Lambda^{(3)} &= \big\{(l_1,l_2,-\wtl_1)~\text{and}~(l_1,-\wtl_2,-\wtl_1)\big\}.
\end{align}
\end{subequations}
\end{addmargin}
%\end{mdframed}
\rule{\linewidth}{0.5pt}

For each subspace \(\mathcal{V}_L\), we focus on the primary operator in the conformal representation, \textit{i.e.} the state in \(R_{[\Delta,j_1,j_2]}\) annihilated by \(K^{a\da}\). These are the states which are harmonics on the Stiefel manifold, and we refer to them as Stiefel harmonics. It is important to recognize that a harmonic in the \(W_L\) representation of \(U(N)\) contains \(\text{dim}(W_L)\) primary operators, and that the different primaries may be composed of particles of different spin (see eq.~\eqref{eq:stress_tensor} for an explicit example). The physical basis for the states in a given harmonic are characterized by the \(U(1)^N\) little group scalings; a prescription to construct these states is supplied by semi-standard Young tableau (SSYT).

%%%%%%%%%%%%%%%%%%%%%%%%%%%%%%%%%%%
%%%%%%%%%%%%%%%%%%%%%%%%%%%%%%%%%%%
%%%%%%%%%%%%%%%%%%%%%%%%%%%%%%%%%%%

\section{ Constructing harmonics}
 \begin{figure}
\includegraphics[width=6.5cm]{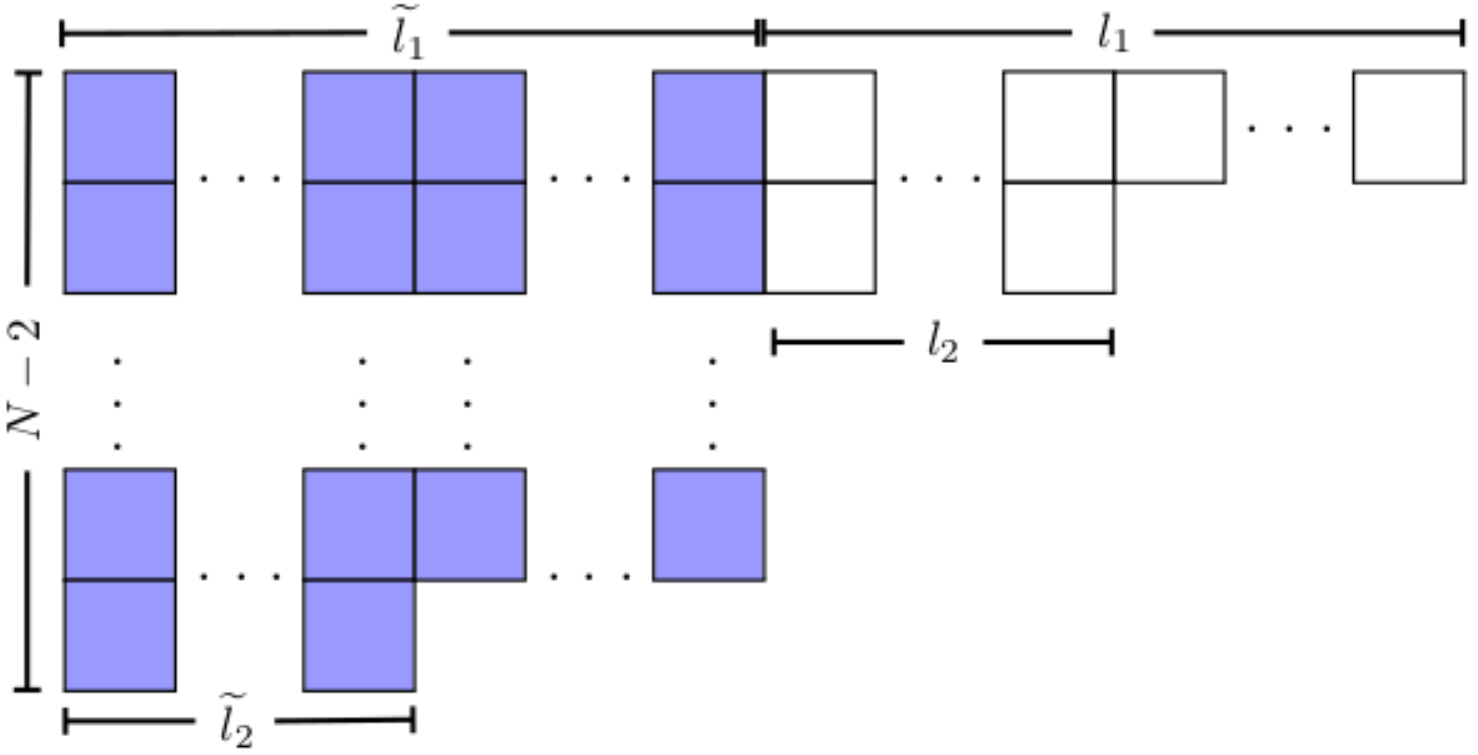}
\caption{
 Young diagram for the $U(N)$ representation $\Lambda^{(N)}$ given in eq.~\eqref{eq:masterweight}.}
 \label{fig:glue}
 \end{figure}

We proceed algebraically by organizing the space of polynomials in the \(N\) spinors into finite-dimensional irreps of the Lorentz group and \(U(N)\) {(we focus for now on the Lorentz subgroup \(SL(2,\mathbb{C}) \subset SU(2,2)\))}. As usual, these are characterized by Young diagrams, which ultimately provide a recipe for symmetrizing indices. 

The $\la$s  are in an irrep of $SL(2,\mathbb{C})\otimes U(N)$,
\be\ytableausetup{boxsize = 1em}
\la^{i}_a = V\otimes W = {\tiny   \begin{array}{c}\begin{ytableau} ~\cr \end{ytableau}\end{array}  \otimes  \begin{array}{c}\begin{ytableau} ~\cr \end{ytableau}\end{array} } \,,
\ee
where $V$ denotes the spinor rep of $SL(2,\mathbb{C})$ and $W = W_{(1,0,\dots,0)}$ the fundamental of $U(N)$. We introduced 
Young diagram notation  (see \textit{e.g.}~\cite{Georgi:1982jb}) to depict both $SL(2,\mathbb{C})$ and $U(N)$ reps  (diagrams with at most two rows and $N$ rows, respectively). For the $\lt$s,
\be\ytableausetup{boxsize = 1em}
\lt_{\da i} = V^*\otimes \overline{W} = {\tiny \begin{array}{c} \begin{ytableau} ~\cr \end{ytableau}\end{array} \otimes \begin{array}{c}\oline{\begin{ytableau} ~\cr \end{ytableau} } \end{array} } =
{\tiny \begin{array}{c} \begin{ytableau} ~\cr \end{ytableau} \end{array} \otimes
 \begin{array}{c}
\begin{ytableau} 
*(blue!40)   \cr *(blue!40)   \cr 
\end{ytableau}\\
\cdot\\\cdot\\
\begin{ytableau} 
 *(blue!40)   \cr
\end{ytableau}
\end{array}}\,,
\ee
where $V^*$ denotes the complex conjugate spinor rep of $SL(2,\mathbb{C})$, and $\overline{W} = W_{(0,\dots,0,-1)}$ the anti-fundamental of $U(N)$.  Young diagrams for conjugate $U(N)$ representations are barred; by using the epsilon tensor to raise 
indices, these can be 
expressed by exchanging each column of $x$ boxes by a column of $N-x$ boxes (and reversing column order
to make a valid diagram). We shade such ``raised'' boxes blue to keep track of their $\lt$ origin.

The simple observation is that for a polynomial built from an object with two indices, \textit{e.g.} \(\la_a^i\), symmetrizing on one set of indices according to some Young diagram automatically symmetrizes the other set of indices according to the same Young diagram.  That is, the organisation of \(\la_{a_1}^{i_1} \cdots \la_{a_n}^{i_n}\) (\(=\text{Sym}^n(\la) = \text{Sym}^n(V \otimes W)\), with Sym denoting the symmetric tensor product) into \(SL(2,\mathbb{C})\times U(N)\) multiplets takes the following form,
\begin{equation}\label{eq:magic}
  \text{Sym}^n(V\otimes W) =\!\!\!\!\!   \!\!\!\!\!   \!\!\!\!\!\!\!\!\!   \!\!\!\!\!   \!\!\!\!\ \bigoplus_{{\footnotesize \begin{array}{c}\rho \vdash n \\ \text{Length}(\rho) \le \text{min}(\text{dim}V,\text{dim}W) \end{array}}}  \!\!\!\!\!   \!\!\!\!\!   \!\!\!\!\!   \!\!\!\!\!   \!\!\!\!\! \mathbf{S}_{\rho}(V)~\otimes~\mathbf{S}_{\rho}(W) \,,
\end{equation}
where $\mathbf{S}_{\rho}$
symmetrizes the indices according to the partition $\rho$ and the cutoff on the number of rows in \(\rho\) is because one can only antisymmetrize so much before an object vanishes. Because dim$V=2$, for $N\ge2$ we have Length$(\rho)\le2$, and
so the sum is over partitions $(l_1,l_2)$ with $l_1+l_2=n$.

Polynomials in the $\la$ and $\lt$ live in a polynomial ring \(R_N = \mathbb{C}[\la,\lt]\) that can be decomposed as 

\be
R_N &= & \bigoplus_{k=0}^{\infty} \text{Sym}^k\big(\la \oplus \lt \big) \nonumber \\
&=& \bigoplus_{k=0}^{\infty}\bigoplus_{n+\nt = k} \text{Sym}^n(\la) \otimes \text{Sym}^{\nt}(\lt) \,, \nonumber \\
&=& \bigoplus_{k=0}^{\infty}\bigoplus_{n+\nt = k} \left\{ \left( \bigoplus_{l_1+l_2=n}   V_{(l_1,l_2)}\otimes W_{(l_1,l_2,0,..,0)}  \right) \nonumber  \right. \\
&~&\left. ~~\otimes \left( \bigoplus_{\widetilde{l}_1+\widetilde{l}_2 = \nt}  V^*_{(\widetilde{l}_1,\widetilde{l}_2)}\otimes {W}_{(0,..,0,-\widetilde{l}_2,-\widetilde{l}_1)}\right) \right\} ,  \,\, \label{eq:ring_decomp}
\ee
where we have used eq.~\eqref{eq:magic} and assummed \(N\ge 2\). While the Lorentz structure in the above is irreducible, the products $\nobreak{W_{{(l_1,l_2,0,..,0)}} \otimes {W}_{{(0,..,0,-\widetilde{l}_2,-\widetilde{l}_1})}}$ are reducible under \(U(N)\). 
  Representing ${W}_{{(0,..,0,-\widetilde{l}_2,-\widetilde{l}_1})}$
  by the  conjugate (blue shaded) diagram, the familiar ``box-placing'' (Littlewood-Richardson)  rules
 for decomposing $U(N)$ tensor products can be applied. We take \(N \ge 4\) for the moment (see below for \(N < 4\)). For every $l_1,l_2,\widetilde{l}_1,\widetilde{l}_2$, only one irrep in this decomposition is primary: the one obtained by simply ``gluing'' the conjugated Young diagram for ${W}_{{(0,..,0,-\widetilde{l}_2,-\widetilde{l}_1})}$ to the left of the diagram  for $W_{{(l_1,l_2,0,..,0)}}$. We show the resulting Young diagram in Fig.~\ref{fig:glue}; it corresponds to the irrep $W_{{(l_1,l_2,0,..,0,-\widetilde{l}_2,-\widetilde{l}_1})}$ of the main result, eq.~\eqref{eq:masterweight}. 
 
 That the other terms in the decomposition of $\nobreak{W_{{(l_1,l_2,0,..,0)}} \otimes {W}_{{(0,..,0,-\widetilde{l}_2,-\widetilde{l}_1})}}$ are not
 primary is readily verified by appealing to the ``box-placing''  rules, from which two situations arise. A white box placed below a column of $N-1$ blue boxes 
 directly factors $\sum_k \lt_k  \lambda^k= P$ in the operator; a white box placed below a column of $N-2$ blue boxes factors terms
 of schematic form $\sum_{k=1}^{N-1} \lt_{N\,\dot{a}} \lt_k^{\dot{a}} \la^{k\,a}=\lt_{N\,\dot{a}}P^{\dot{a}a}$. The presence of
 a factor of $P$ indicates the operator is a descendant and ensures the non-annihilation by $K$.

 Because the primary operators are in the \(U(N)\) irrep \((l_1,l_2,0,\dots,0,-\wtl_2,-\wtl_1)\), which is simply the partition obtained by adding \((l_1,l_2,0,\dots)\) and \((\dots,0,-\wtl_2,-\wtl_1)\), for \(N<4\) this obviously needs further consideration. Descendants contain overall factors of \(P\); these get factored out of the polynomial \(W_{(l_1,l_2,0,\dots)}\otimes W_{(\dots,0,-\wtl_2,-\wtl_1)}\) when an upper \(\la^i\) index contracts a lower \(\lt_i\) index. For \(N<4\) this happens if numbers in the added partition cancel; \textit{e.g.} for \(N=3\), if both \(l_2,\wtl_2 \ne 0\) in \((l_1,l_2 - \wtl_2,-\wtl_1)\). This rule readily gives eq.~\eqref{eq:exceptional}. An important consequence is that scattering operators---Lorentz invariant operators with derivatives acting on fields (\(l_1 = l_2 \ne 0\) and \(\wtl_1 = \wtl_2 \ne 0\))---only occur for \(N\ge 4\); this is the familiar statement that Mandelstam invariants are only non-trivial for \(N \ge 4\).

The states of the $U(N)$ rep can be constructed using SSYT: fillings of the Young diagram boxes with numbers from the set $1,\ldots,N$,
such that numbers weakly increase across rows and strongly increase down columns. The number of valid SSYT is equal to the dimension of the $U(N)$ rep. The highest weight state fills the $k$th row with the number $k$. One can verify that this state is trivially annihilated by $K$; that all other states in the rep are too follows because $K$ is a $U(N)$ singlet.

\section{Example: \(N=2\) \& higher spin currents}

Consider the \(N=2\) sector \(\scH_2\). The holomorphic---built only from \(\la\)s---primary operators belong to the \(U(2)\) representations \((l_1,l_2)\) (likewise, the anti-holomorphic primaries fall into the \((-\wtl_2,-\wtl_1)\) representations). When \(l_1 = l_2 \equiv l\) the harmonics are Lorentz scalars; moreover, they contain only a single operator (they are singlets under the \(SU(2) \subset U(2)\)). For \(l\ge 2\) these correspond to squared field-strength terms for helicity \(l/2\) particles
\be
\ytableausetup{boxsize = 1em}
  \underbrace{\begin{ytableau}  1 & 1 \\ 2 & 2 \end{ytableau} \ldots \begin{ytableau} 1  \\ 2 \end{ytableau}}_l ~ \Leftrightarrow ~ (\la_{a}^1 \la^{2\,a})^l \sim (F_{1,L})_{a_1\dots a_l} (F_{2,L})^{a_1 \dots a_l} .  \nonumber
\ee
Note that to describe indistinguishable particles 1 and 2, we must symmetrize over the indices $(1,2)$ for the bosonic (even $l$) case, or else anti-symmetrize for the fermionic (odd $l$) case. In both cases, the operator does not vanish.

The non-holomorphic harmonics are particularly interesting for \(N=2\), as they correspond to conserved currents like the stress tensor and higher spin currents. Here we describe them and provide a generating function; further details can be found in the supplemental material.

The non-holomorphic harmonics in \(\scH_2\) are the \(U(2)\) representations
\be
\ytableausetup{boxsize = .7em}
  (n,-m) \Leftrightarrow \underbrace{\begin{ytableau} *(blue!40)  \end{ytableau} \cdots \begin{ytableau} *(blue!40) \end{ytableau}}_{m} \hspace{-1.5pt} \underbrace{\begin{ytableau} ~ \end{ytableau} \cdots \begin{ytableau} ~ \end{ytableau}}_{n} 
\label{eq:HS_YD}
\ee
with \(n,m >0\) and carrying spin \((j_1,j_2) = (\frac{n}{2},\frac{m}{2})\). Recall that the conjugated diagrams involve raising the \(U(2)\) index on the \(\lt\)s,
\begin{equation}\ytableausetup{boxsize = .7em}
  \overline{\ydiagram{1}} \to \begin{ytableau} *(blue!40)   \cr \end{ytableau} ~~ \Leftrightarrow ~~ \lt_{i\da} \to \lt^i_{\da} = \epsilon^{ij}\lt_{j\da},
\end{equation}
so that the polynomial encoded by eq.~\eqref{eq:HS_YD} simply symmetrizes over all of the (raised) flavor indices. We define this operator
\begin{align}
  J^{(n,m)} &= \big(J^{(n,m)}\big)^{(i_1\dots i_{n+m})}_{(a_1\dots a_n)(\da_1\dots \da_m)} \nonumber \\
  &= \la^{(i_1}_{a_1}\cdots \la^{i_n}_{a_n} \lt^{i_{n+1}}_{\da_1} \cdots \lt^{i_{n+m})}_{\da_{m}}.
\end{align}
There are \(n+m+1\) primaries inside this \(U(2)\) representation, corresponding to setting \((i_1,\ldots,i_{n+m}) = (1,\ldots,1),\ (1,\dots,1,2),\, \ldots \,, (1,2,\dots,2), \text{ or } (2,\dots,2)\). For shorthand, we denote these states by
\be
  J^{(n,m)}_{1^{n+m-k}2^k} \equiv \big(J^{(n,m)}\big)_{(a_1 \dots a_n)(\da_1 \dots \da_m)}^{(\overbrace{1 \dots 1}^{n+m-k}\overbrace{2\dots 2}^k)}.
\ee
We can readily come up with a generating function for these states. Let \(\la^1_a \equiv \la_a\) and \(\la^2_a \equiv \et_a\) and define
\be
  f^{(n,m)} \equiv \prod_{i=1}^n\big(c \la + \et\big )_{a_i} \prod_{j=1}^m \big( \etl - \tfrac{1}{c} \lt\big)_{\da_j},
\ee
with \(c \in \mathbb{C}\) some arbitrary parameter. The chain rule readily verifies that \(K^{a\da} = -(\pd_{\la}\pdt_{\la} + \pd_{\et}\pdt_{\et})^{a\da}\) annihilates \(f^{(n,m)}\). Since \(c\) is arbitrary, this further implies that each term in the expansion \(f^{(n,m)} = \sum_{k=0}^{n+m} c^{n-k}f^{(n,m)}_k\) is also annihilated by \(K\). With a little more effort, one deduces
\be
f^{(n,m)} = \sum_{k=0}^{n+m} c^{n-k} \binom{n+m}{k} J^{(n,m)}_{1^{n+m-k}2^k},
\label{eq:fnm_expansion}
\ee
thereby providing an efficient means to obtain the \(J^{(n,m)}_{1^{n+m-k}2^k}\). Finally, note that \(P\cdot f^{(n,m)} \equiv P^{a \da} f^{(n,m)}_{a a_2\dots a_n \da \da_2 \dots \da_m} = 0\); this implies \(P\cdot J^{(n,m)}_{1^{n+m-k}2^k} = 0\), \textit{i.e.} these are conserved currents.

As an example, consider the stress-tensor harmonic \(n=m=2 \rightarrow (j_1,j_2) = (1,1)\). The five states in \(J^{(2,2)}\) are 
\be \label{eq:stress_tensor}
 {\tiny  J^{(2,2)}} ={ \left\{ \def\arraystretch{1.5} {\small \begin{array}{l}
        (\la\la\etl\etl)_{ab\da\db} \\
      \frac{1}{4}(-\la\la\etl\lt - \la\la\lt\etl + \la\et\etl\etl +\et \la\etl\etl)_{ab\da\db} \\
      \frac{1}{6}(\la\la\lt\lt - \la\et\etl\lt - \et\la\etl\lt - \la\et\lt\etl - \et\la\lt\etl + \et\et\etl\etl)_{ab\da\db}\\
      \frac{1}{4}(\la\et\lt\lt + \et\la\lt\lt - \et\et\etl\lt - \et\et\lt\etl)_{ab\da\db} \\
      (\et\et\lt\lt)_{ab\da\db} 
  \end{array} } \right.}  ~~~~
\ee
corresponding to operators of the schematic form $F_{1\,L} F_{2\,R}$, $\psi_{1\,L} \partial \psi_{2\,R}$, $\phi_1\partial \partial \phi_2$, $\psi_{1\,R} \partial \psi_{2\,L}$, and  $F_{1\,R} F_{2\,L}$.

There is something remarkable about this result. These operators carry spin \((j_1,j_2)=(1,1)\), so they could in principle have nine independent components. However, they are conserved, \(P\cdot J^{(2,2)}_{1^{4-k}2^k} = 0\), reducing the number of independent components to five. This is precisely the dimension of the \(U(2)\) representation. In fact, as a result of conservation, this happens for \textit{all} \((n,m)\): the number of independent components of \(J^{(n,m)}_{1^{n+m-k}2^k}\) is equal to \(n+m+1\). Mathematically, this follows from the reductive dual pair structure and is a consequence of Frobenius reciprocity, on which we elaborate further in the supplementary material.

\section{Discussion}
A very useful extension of the results we present above would be a similar systematic understanding of identical particles: an operator containing identical (fermions) bosons lies within the appropriately (anti-)symmetrized Fock space $\subset \mathcal{H}_N$. In practice, such an operator is selected out by applying the permutation group to the ($N$-distinguishable) operators whose construction we detailed. This operation  preserves the primary condition but mixes states within a $U(N)$ representation (in $d=4$ the permutation belongs to the Weyl group  $S_N\subset U(N)$). In this vein, the works~\cite{deMelloKoch:2017caf,deMelloKoch:2017dgi,deMelloKoch:2018fze,deMelloKoch:2018klm} could be useful.

We have focused on the spectrum of free CFTs, and seen how a generalized notion of helicity encodes this information. The other data in CFTs are the OPE coefficients. What, if anything, does the \(U(N)\) (or \(O(N)\) in \(d=2,3\)) say about these? Could it be that the OPE coefficients are related in some way to Clebsch-Gordan coefficients of \(U(N)\)? In this spirit---and discussed briefly in the supplementary material---we note that the use of spinors opens up potentially more efficient methods for evaluating OPE coefficients (as compared to traditional momentum variables).

A supersymmetric version of the oscillator representation  is obtained with the inclusion of anticommuting counterparts to the spinors (see {\it e.g.}~\cite{Witten:2003nn} for such a representation for $\mathcal{N}=4$ SYM); it would be interesting to flesh out the reductive dual pair part of this story.

Underlying the conformal-helicity duality/reductive dual pair structure described in this work is the symplectic action on spaces of polynomials, \textit{i.e.} on the oscillator representation. We have utilized this for finite \(N\), in particular to decompose \(\mathcal{H}_N\) into irreps of the conformal group. Important in this regard where how the \(SU(2,2)\) and \(U(N)\) generators were quadratic in the spinors and their derivatives. The question naturally arises about an infinite dimensional generalization where the oscillator representation arises as automorphisms of the \textit{field theory} canonical commutation relations, \([\phi(\mathbf{x}),\pi(\mathbf{y})] = i \delta(\mathbf{x} - \mathbf{y})\). (In fact, this infinite dimensional case is where the oscillator representation was formally introduced by Segal and Shale~\cite{segal1963mathematical, shale1962linear}.) We expect the higher spin currents (which are quadratic in the fields) to play an important role; it would be interesting to see if this allows a more efficient means to constructing the spectrum and the OPE coefficients of free theories.

What more of the $U(N)$? Certainly its representation theory is relevant to the discussion of non-renormalisation theories in EFT~\cite{Alonso:2014rga,Cheung:2015aba,Elias-Miro:2014eia}, where the grouping of operators~\cite{Cheung:2015aba} corresponds to those that belong to the same $U(N)$ representation.   And we have seen above how the $U(2)$ encodes properties of higher spin currents at a deep mathematical level; it would be nice to gain a physical understanding of this, if indeed such an understanding exists.

\section*{Acknowledgements}
We are grateful for conversations with Walter Goldberger, Simeon Hellerman, Mikhail Kapranov, Xiaochuan Lu, Hitoshi Murayama, David Poland, Siddarth Prabhu, Witek Skiba, Matt Walters, and Junpu Wang. We thank Julian Sonner for comments on a draft of this work. We additionally thank the organizers and participants at the IHES workshop on ``Hamiltonian methods in strongly coupled field theory''. We extend a special thank you to Jed Thompson and Gregg Zuckerman for many delightful and enlightening conversations. BH is funded by the Swiss National Science Foundation under grant no. PP002-170578. TM is supported by the World Premier International Research Center Initiative (WPI), MEXT, Japan, and by JSPS KAKENHI Grant Number JP18K13533.

\appendix

\bibliographystyle{utphys}
\bibliography{bibliography}

\providecommand{\href}[2]{#2}\begingroup\raggedright\begin{thebibliography}{10}

\bibitem{Froissart}
M.~Froissart, ``{Asymptotic Behavior and Subtractions in the Mandelstam
  Representation},''
{\em Phys. Rev.} {\bf 123} (1961)  1053--1057.
%%CITATION = JTPLA,41,667;%%.

\bibitem{Gribov:1961ex}
V.~N. Gribov, ``{Possible Asymptotic Behavior of Elastic Scattering},''
{\em JETP Lett.} {\bf 41} (1961)  667--669.
%%CITATION = JTPLA,41,667;%%.

\bibitem{Adams:2006sv}
A.~Adams, N.~Arkani-Hamed, S.~Dubovsky, A.~Nicolis, and R.~Rattazzi,
  ``{Causality, analyticity and an IR obstruction to UV completion},''
  \href{http://dx.doi.org/10.1088/1126-6708/2006/10/014}{{\em JHEP} {\bf 10}
  (2006)  014},
\href{http://arxiv.org/abs/hep-th/0602178}{{\tt arXiv:hep-th/0602178
  [hep-th]}}.
%%CITATION = HEP-TH/0602178;%%.

\bibitem{Paulos:2017fhb}
M.~F. Paulos, J.~Penedones, J.~Toledo, B.~C. van Rees, and P.~Vieira, ``{The
  S-matrix Bootstrap III: Higher Dimensional Amplitudes},''
\href{http://arxiv.org/abs/1708.06765}{{\tt arXiv:1708.06765 [hep-th]}}.
%%CITATION = ARXIV:1708.06765;%%.

\bibitem{Guerrieri:2018uew}
A.~L. Guerrieri, J.~Penedones, and P.~Vieira, ``{Bootstrapping QCD: the Lake,
  the Peninsula and the Kink},''
\href{http://arxiv.org/abs/1810.12849}{{\tt arXiv:1810.12849 [hep-th]}}.
%%CITATION = ARXIV:1810.12849;%%.

\bibitem{Hogervorst:2014rta}
M.~Hogervorst, S.~Rychkov, and B.~C. van Rees, ``{Truncated conformal space
  approach in d dimensions: A cheap alternative to lattice field theory?},''
  \href{http://dx.doi.org/10.1103/PhysRevD.91.025005}{{\em Phys. Rev.} {\bf
  D91} (2015)  025005},
\href{http://arxiv.org/abs/1409.1581}{{\tt arXiv:1409.1581 [hep-th]}}.
%%CITATION = ARXIV:1409.1581;%%.

\bibitem{Katz:2016hxp}
E.~Katz, Z.~U. Khandker, and M.~T. Walters, ``{A Conformal Truncation Framework
  for Infinite-Volume Dynamics},''
  \href{http://dx.doi.org/10.1007/JHEP07(2016)140}{{\em JHEP} {\bf 07} (2016)
  140},
\href{http://arxiv.org/abs/1604.01766}{{\tt arXiv:1604.01766 [hep-th]}}.
%%CITATION = ARXIV:1604.01766;%%.

\bibitem{Katz:2013qua}
E.~Katz, G.~Marques~Tavares, and Y.~Xu, ``{Solving 2D QCD with an adjoint
  fermion analytically},''
  \href{http://dx.doi.org/10.1007/JHEP05(2014)143}{{\em JHEP} {\bf 05} (2014)
  143},
\href{http://arxiv.org/abs/1308.4980}{{\tt arXiv:1308.4980 [hep-th]}}.
%%CITATION = ARXIV:1308.4980;%%.

\bibitem{Katz:2014uoa}
E.~Katz, G.~Marques~Tavares, and Y.~Xu, ``{A solution of 2D QCD at Finite $N$
  using a conformal basis},''
\href{http://arxiv.org/abs/1405.6727}{{\tt arXiv:1405.6727 [hep-th]}}.
%%CITATION = ARXIV:1405.6727;%%.

\bibitem{Rychkov:2014eea}
S.~Rychkov and L.~G. Vitale, ``{Hamiltonian truncation study of the $\phi^4$
  theory in two dimensions},''
  \href{http://dx.doi.org/10.1103/PhysRevD.91.085011}{{\em Phys. Rev.} {\bf
  D91} (2015)  085011},
\href{http://arxiv.org/abs/1412.3460}{{\tt arXiv:1412.3460 [hep-th]}}.
%%CITATION = ARXIV:1412.3460;%%.

\bibitem{Anand:2017yij}
N.~Anand, V.~X. Genest, E.~Katz, Z.~U. Khandker, and M.~T. Walters, ``{RG flow
  from $\phi^4$ theory to the 2D Ising model},''
  \href{http://dx.doi.org/10.1007/JHEP08(2017)056}{{\em JHEP} {\bf 08} (2017)
  056},
\href{http://arxiv.org/abs/1704.04500}{{\tt arXiv:1704.04500 [hep-th]}}.
%%CITATION = ARXIV:1704.04500;%%.

\bibitem{Delacretaz:2018xbn}
L.~V. Delacrétaz, A.~L. Fitzpatrick, E.~Katz, and L.~G. Vitale, ``{Conformal
  Truncation of Chern-Simons Theory at Large $N_f$},''
\href{http://arxiv.org/abs/1811.10612}{{\tt arXiv:1811.10612 [hep-th]}}.
%%CITATION = ARXIV:1811.10612;%%.

\bibitem{Henning:2015daa}
B.~Henning, X.~Lu, T.~Melia, and H.~Murayama, ``{Hilbert series and operator
  bases with derivatives in effective field theories},''
  \href{http://dx.doi.org/10.1007/s00220-015-2518-2}{{\em Commun. Math. Phys.}
  {\bf 347} (2016) no.~2, 363--388},
\href{http://arxiv.org/abs/1507.07240}{{\tt arXiv:1507.07240 [hep-th]}}.
%%CITATION = ARXIV:1507.07240;%%.

\bibitem{Henning:2015alf}
B.~Henning, X.~Lu, T.~Melia, and H.~Murayama, ``{2, 84, 30, 993, 560, 15456,
  11962, 261485, ...: Higher dimension operators in the SM EFT},''
  \href{http://dx.doi.org/10.1007/JHEP08(2017)016}{{\em JHEP} {\bf 08} (2017)
  016},
\href{http://arxiv.org/abs/1512.03433}{{\tt arXiv:1512.03433 [hep-ph]}}.
%%CITATION = ARXIV:1512.03433;%%.

\bibitem{Henning:2017fpj}
B.~Henning, X.~Lu, T.~Melia, and H.~Murayama, ``{Operator bases, $S$-matrices,
  and their partition functions},''
  \href{http://dx.doi.org/10.1007/JHEP10(2017)199}{{\em JHEP} {\bf 10} (2017)
  199},
\href{http://arxiv.org/abs/1706.08520}{{\tt arXiv:1706.08520 [hep-th]}}.
%%CITATION = ARXIV:1706.08520;%%.

\bibitem{Lehman:2015via}
L.~Lehman and A.~Martin, ``{Hilbert Series for Constructing Lagrangians:
  expanding the phenomenologist's toolbox},''
  \href{http://dx.doi.org/10.1103/PhysRevD.91.105014}{{\em Phys. Rev.} {\bf
  D91} (2015)  105014},
\href{http://arxiv.org/abs/1503.07537}{{\tt arXiv:1503.07537 [hep-ph]}}.
%%CITATION = ARXIV:1503.07537;%%.

\bibitem{Lehman:2015coa}
L.~Lehman and A.~Martin, ``{Low-derivative operators of the Standard Model
  effective field theory via Hilbert series methods},''
  \href{http://dx.doi.org/10.1007/JHEP02(2016)081}{{\em JHEP} {\bf 02} (2016)
  081},
\href{http://arxiv.org/abs/1510.00372}{{\tt arXiv:1510.00372 [hep-ph]}}.
%%CITATION = ARXIV:1510.00372;%%.

\bibitem{Craigie:1983fb}
N.~S. Craigie, V.~K. Dobrev, and I.~T. Todorov, ``{Conformally Covariant
  Composite Operators in Quantum Chromodynamics},''
\href{http://dx.doi.org/10.1016/0003-4916(85)90118-6}{{\em Annals Phys.} {\bf
  159} (1985)  411--444}.
%%CITATION = APNYA,159,411;%%.

\bibitem{Braun:2003rp}
V.~M. Braun, G.~P. Korchemsky, and D.~Mueller, ``{The Uses of conformal
  symmetry in QCD},''
  \href{http://dx.doi.org/10.1016/S0146-6410(03)90004-4}{{\em Prog. Part. Nucl.
  Phys.} {\bf 51} (2003)  311--398},
\href{http://arxiv.org/abs/hep-ph/0306057}{{\tt arXiv:hep-ph/0306057
  [hep-ph]}}.
%%CITATION = HEP-PH/0306057;%%.

\bibitem{Giombi:2016hkj}
S.~Giombi and V.~Kirilin, ``{Anomalous dimensions in CFT with weakly broken
  higher spin symmetry},''
  \href{http://dx.doi.org/10.1007/JHEP11(2016)068}{{\em JHEP} {\bf 11} (2016)
  068},
\href{http://arxiv.org/abs/1601.01310}{{\tt arXiv:1601.01310 [hep-th]}}.
%%CITATION = ARXIV:1601.01310;%%.

\bibitem{deMelloKoch:2017caf}
R.~de~Mello~Koch, P.~Rabambi, R.~Rabe, and S.~Ramgoolam, ``{Free quantum fields
  in 4D and Calabi-Yau spaces},''
  \href{http://dx.doi.org/10.1103/PhysRevLett.119.161602}{{\em Phys. Rev.
  Lett.} {\bf 119} (2017) no.~16, 161602},
\href{http://arxiv.org/abs/1705.04039}{{\tt arXiv:1705.04039 [hep-th]}}.
%%CITATION = ARXIV:1705.04039;%%.

\bibitem{companion}
B.~Henning and T.~Melia, ``{Constructing effective field theories via their
  harmonics},''
\href{http://arxiv.org/abs/1902.xxxxx}{{\tt arXiv:1902.xxxxx [hep-ph]}}.
%%CITATION = ARXIV:1706.02712;%%.

\bibitem{siegel}
W.~SIEGEL, ``All free conformal representations in all dimensions,''
  \href{http://dx.doi.org/10.1142/S0217751X89000819}{{\em International Journal
  of Modern Physics A} {\bf 04} (1989) no.~08, 2015--2020}.

\bibitem{ArkaniHamed:2009dn}
N.~Arkani-Hamed, F.~Cachazo, C.~Cheung, and J.~Kaplan, ``{A Duality For The S
  Matrix},'' \href{http://dx.doi.org/10.1007/JHEP03(2010)020}{{\em JHEP} {\bf
  03} (2010)  020},
\href{http://arxiv.org/abs/0907.5418}{{\tt arXiv:0907.5418 [hep-th]}}.
%%CITATION = ARXIV:0907.5418;%%.

\bibitem{Kravchuk:2016qvl}
P.~Kravchuk and D.~Simmons-Duffin, ``{Counting Conformal Correlators},''
  \href{http://dx.doi.org/10.1007/JHEP02(2018)096}{{\em JHEP} {\bf 02} (2018)
  096},
\href{http://arxiv.org/abs/1612.08987}{{\tt arXiv:1612.08987 [hep-th]}}.
%%CITATION = ARXIV:1612.08987;%%.

\bibitem{howe}
R.~Howe, ``Remarks on classical invariant theory,'' {\em Transactions of the
  American Mathematical Society} {\bf 313} (1989) no.~2, 539--570.
  \url{http://www.jstor.org/stable/2001418}.

\bibitem{howe1985}
R.~Howe, ``Dual pairs in physics: harmonic oscillators, photons, electrons, and
  singletons,'' {\em Lectures in Applied Math.} {\bf 21} (1985)  179--207.

\bibitem{MackTod}
G.~Mack and I.~Todorov, ``{Irreducibility of the ladder representations of
  u(2,2) when restricted to the poincare subgroup},''
\href{http://dx.doi.org/10.1063/1.1664804}{{\em J. Math. Phys.} {\bf 10} (1969)
   2078--2085}.
%%CITATION = JMAPA,10,2078;%%.

\bibitem{kashiwara1978}
M.~Kashiwara and M.~Vergne, ``On the segal-shale-weil representations and
  harmonic polynomials,'' {\em Inventiones mathematicae} {\bf 44} (1978) no.~1,
  1--47.

\bibitem{gelbart1974}
S.~S. Gelbart, ``A theory of stiefel harmonics,'' {\em Transactions of the
  American Mathematical Society} {\bf 192} (1974)  29--50.

\bibitem{Mack:1975}
G.~Mack, ``{All unitary ray representations of the conformal group SU(2,2) with
  positive energy},''
\href{http://dx.doi.org/10.1007/BF01613145}{{\em Commun. Math. Phys.} {\bf 55}
  (1977)  1}.
%%CITATION = CMPHA,55,1;%%.

\bibitem{Fulton}
W.~Fulton and J.~Harris, ``{Representation Theory: A First Course},'' {\em
  Springer} (1991)  .

\bibitem{Georgi:1982jb}
H.~Georgi, ``{LIE ALGEBRAS IN PARTICLE PHYSICS. FROM ISOSPIN TO UNIFIED
  THEORIES},''
{\em Front. Phys.} {\bf 54} (1982)  1--255.
%%CITATION = FRPHA,54,1;%%.

\bibitem{deMelloKoch:2017dgi}
R.~de~Mello~Koch, P.~Rabambi, R.~Rabe, and S.~Ramgoolam, ``{Counting and
  construction of holomorphic primary fields in free CFT4 from rings of
  functions on Calabi-Yau orbifolds},''
  \href{http://dx.doi.org/10.1007/JHEP08(2017)077}{{\em JHEP} {\bf 08} (2017)
  077},
\href{http://arxiv.org/abs/1705.06702}{{\tt arXiv:1705.06702 [hep-th]}}.
%%CITATION = ARXIV:1705.06702;%%.

\bibitem{deMelloKoch:2018fze}
R.~De~Mello~Koch, P.~Rambambi, and H.~J.~R. Van~Zyl, ``{From Spinning Primaries
  to Permutation Orbifolds},''
  \href{http://dx.doi.org/10.1007/JHEP04(2018)104}{{\em JHEP} {\bf 04} (2018)
  104},
\href{http://arxiv.org/abs/1801.10313}{{\tt arXiv:1801.10313 [hep-th]}}.
%%CITATION = ARXIV:1801.10313;%%.

\bibitem{deMelloKoch:2018klm}
R.~de~Mello~Koch and S.~Ramgoolam, ``{Free field primaries in general
  dimensions: Counting and construction with rings and modules},''
  \href{http://dx.doi.org/10.1007/JHEP08(2018)088}{{\em JHEP} {\bf 08} (2018)
  088},
\href{http://arxiv.org/abs/1806.01085}{{\tt arXiv:1806.01085 [hep-th]}}.
%%CITATION = ARXIV:1806.01085;%%.

\bibitem{Witten:2003nn}
E.~Witten, ``{Perturbative gauge theory as a string theory in twistor space},''
  \href{http://dx.doi.org/10.1007/s00220-004-1187-3}{{\em Commun. Math. Phys.}
  {\bf 252} (2004)  189--258},
\href{http://arxiv.org/abs/hep-th/0312171}{{\tt arXiv:hep-th/0312171
  [hep-th]}}.
%%CITATION = HEP-TH/0312171;%%.

\bibitem{segal1963mathematical}
I.~E. Segal, {\em Mathematical problems of relativistic physics}, vol.~2.
\newblock American Mathematical Soc., 1963.

\bibitem{shale1962linear}
D.~Shale, ``Linear symmetries of free boson fields,'' {\em Transactions of the
  American Mathematical Society} {\bf 103} (1962) no.~1, 149--167.

\bibitem{Alonso:2014rga}
R.~Alonso, E.~E. Jenkins, and A.~V. Manohar, ``{Holomorphy without
  Supersymmetry in the Standard Model Effective Field Theory},''
  \href{http://dx.doi.org/10.1016/j.physletb.2014.10.045}{{\em Phys. Lett.}
  {\bf B739} (2014)  95--98},
\href{http://arxiv.org/abs/1409.0868}{{\tt arXiv:1409.0868 [hep-ph]}}.
%%CITATION = ARXIV:1409.0868;%%.

\bibitem{Cheung:2015aba}
C.~Cheung and C.-H. Shen, ``{Nonrenormalization Theorems without
  Supersymmetry},''
  \href{http://dx.doi.org/10.1103/PhysRevLett.115.071601}{{\em Phys. Rev.
  Lett.} {\bf 115} (2015) no.~7, 071601},
\href{http://arxiv.org/abs/1505.01844}{{\tt arXiv:1505.01844 [hep-ph]}}.
%%CITATION = ARXIV:1505.01844;%%.

\bibitem{Elias-Miro:2014eia}
J.~Elias-Miro, J.~R. Espinosa, and A.~Pomarol, ``{One-loop non-renormalization
  results in EFTs},''
  \href{http://dx.doi.org/10.1016/j.physletb.2015.05.056}{{\em Phys. Lett.}
  {\bf B747} (2015)  272--280},
\href{http://arxiv.org/abs/1412.7151}{{\tt arXiv:1412.7151 [hep-ph]}}.
%%CITATION = ARXIV:1412.7151;%%.

\bibitem{howe2012}
R.~E. Howe and E.~C. Tan, {\em Non-Abelian Harmonic Analysis: Applications of
  $SL(2,\mathbb{R})$}.
\newblock Springer, 2012.

\bibitem{Bargmann:1970}
V.~Bargmann,
``{Group representations on hilbert spaces of analytic functions},''.
%%CITATION = INSPIRE-63677;%%.

\bibitem{Woit:2017}
P.~Woit, \href{http://dx.doi.org/10.1007/978-3-319-64612-1}{{\em {Quantum
  Theory, Groups and Representations}}}.
\newblock Springer, 2017.
\newblock
\url{http://inference-review.com/article/woits-way}.
\newblock
%%CITATION = INSPIRE-1678003;%%.

\bibitem{Luscher:1974}
{L\"uscher, M. and Mack, G.}, ``{Global Conformal Invariance in Quantum Field
  Theory},''
\href{http://dx.doi.org/10.1007/BF01608988}{{\em Commun. Math. Phys.} {\bf 41}
  (1975)  203--234}.
%%CITATION = CMPHA,41,203;%%.

\bibitem{MackSalam}
G.~Mack and A.~Salam, ``{Finite component field representations of the
  conformal group},''
\href{http://dx.doi.org/10.1016/0003-4916(69)90278-4}{{\em Annals Phys.} {\bf
  53} (1969)  174--202}.
%%CITATION = APNYA,53,174;%%.

\bibitem{Ruehl:1973}
W.~Ruehl, ``{Field representations of the conformal group with continuous mass
  spectrum},''
\href{http://dx.doi.org/10.1007/BF01645506}{{\em Commun. Math. Phys.} {\bf 30}
  (1973)  287--302}.
%%CITATION = CMPHA,30,287;%%.

\bibitem{Ruehl:1973pr}
{R\"uhl, W.}, ``{On conformal invariance of interacting fields},''
\href{http://dx.doi.org/10.1007/BF01646444}{{\em Commun. Math. Phys.} {\bf 34}
  (1973)  149--166}.
%%CITATION = CMPHA,34,149;%%.

\bibitem{Bekaert:2006py}
X.~Bekaert and N.~Boulanger, ``{The Unitary representations of the Poincare
  group in any spacetime dimension},'' in {\em {2nd Modave Summer School in
  Theoretical Physics Modave, Belgium, August 6-12, 2006}}.
\newblock 2006.
\newblock
\href{http://arxiv.org/abs/hep-th/0611263}{{\tt arXiv:hep-th/0611263
  [hep-th]}}.
\newblock
%%CITATION = HEP-TH/0611263;%%.

\bibitem{Rychkov:2016}
S.~Rychkov, \href{http://dx.doi.org/10.1007/978-3-319-43626-5}{{\em {EPFL
  Lectures on Conformal Field Theory in $D\ge 3$ Dimensions}}}.
\newblock SpringerBriefs in Physics. 2016.
\newblock
\href{http://arxiv.org/abs/1601.05000}{{\tt arXiv:1601.05000 [hep-th]}}.
\newblock
%%CITATION = ARXIV:1601.05000;%%.

\bibitem{Gillioz:2016}
M.~Gillioz, X.~Lu, and M.~A. Luty, ``{Scale Anomalies, States, and Rates in
  Conformal Field Theory},''
  \href{http://dx.doi.org/10.1007/JHEP04(2017)171}{{\em JHEP} {\bf 04} (2017)
  171},
\href{http://arxiv.org/abs/1612.07800}{{\tt arXiv:1612.07800 [hep-th]}}.
%%CITATION = ARXIV:1612.07800;%%.

\bibitem{streater}
R.~F.~S. und A~S~Wightman, {\em PCT, Spin Statistics, And All That}.
\newblock W A Benjamin Inc., New York, Amsterdam, 1964.

\bibitem{Simmons-Duffin:2016gjk}
D.~Simmons-Duffin, \href{http://dx.doi.org/10.1142/9789813149441_0001}{``{The
  Conformal Bootstrap},''}
\newblock 2017.
\newblock
\href{http://arxiv.org/abs/1602.07982}{{\tt arXiv:1602.07982 [hep-th]}}.
\newblock
%%CITATION = ARXIV:1602.07982;%%.

\bibitem{Coriano:2013}
C.~Coriano, L.~Delle~Rose, E.~Mottola, and M.~Serino, ``{Solving the Conformal
  Constraints for Scalar Operators in Momentum Space and the Evaluation of
  Feynman's Master Integrals},''
  \href{http://dx.doi.org/10.1007/JHEP07(2013)011}{{\em JHEP} {\bf 07} (2013)
  011},
\href{http://arxiv.org/abs/1304.6944}{{\tt arXiv:1304.6944 [hep-th]}}.
%%CITATION = ARXIV:1304.6944;%%.

\bibitem{Bzowski:2013}
A.~Bzowski, P.~McFadden, and K.~Skenderis, ``{Implications of conformal
  invariance in momentum space},''
  \href{http://dx.doi.org/10.1007/JHEP03(2014)111}{{\em JHEP} {\bf 03} (2014)
  111},
\href{http://arxiv.org/abs/1304.7760}{{\tt arXiv:1304.7760 [hep-th]}}.
%%CITATION = ARXIV:1304.7760;%%.

\bibitem{Costa:2011mg}
M.~S. Costa, J.~Penedones, D.~Poland, and S.~Rychkov, ``{Spinning Conformal
  Correlators},'' \href{http://dx.doi.org/10.1007/JHEP11(2011)071}{{\em JHEP}
  {\bf 11} (2011)  071},
\href{http://arxiv.org/abs/1107.3554}{{\tt arXiv:1107.3554 [hep-th]}}.
%%CITATION = ARXIV:1107.3554;%%.

\end{thebibliography}\endgroup

  \newpage
  
\section{Supplementary Material}

\section{\(d=2\) and an analogy to spherical harmonics}
There is a useful analogy between harmonic analysis on the Stiefel manifold and the much more familiar case of harmonic analysis of a sphere. In the latter, functions \(F(x_1,\dots,x_N)\), subject to a constraint on the Cartesian coordinates \(\sum_i x_i^2 = 1\) can be expanded in spherical harmonics, \(F = \sum_{l}\sum_{m_1,\dots,m_{\lfloor \frac{N}{2} \rfloor}}c^l_{\mathbf{m}} Y^l_{\mathbf{m}}\). Alternatively, one can Taylor expand \(F(x_i) = \sum_{l=0}^{\infty}c^{i_1\dots i_l} x_{i_1}\dots x_{i_l}\); as a result of the constraint \(\mathbf{x}^2 = 1\), one needs only take the \textit{traceless} combination \(x_{\{i_1}\dots x_{i_l\}}\) as basis terms in the Taylor expansion. It is not difficult to show that homogeneous, traceless polynomials are harmonic polynomials; that is, they are annihilated by the Laplacian \(\nabla^2 = \sum_i \frac{\pd}{\pd x_i} \frac{\pd}{\pd x_i}\). (Intuitively, one thinks of the Laplacian as enacting a contraction, \(\nabla^2 x_i x_j \propto \delta_{ij}\).) Indeed, by passing to spherical coordinates---\(\nabla^2 = \frac{1}{r^{N-1}}\pd_r r^{N-1} \pd_r + \frac{1}{r^2}\Delta_{S^{N-1}}\) with \(\Delta_{S^{N-1}}\) the Laplacian on the \((N-1)\)-sphere---one immediately sees that a harmonic polynomial restricted to the sphere is an eigenfunction of \(\Delta_{S^{N-1}}\) with eigenvalue \(l(l+N-2)\).

The representation theoretic understanding of the above comes from the natural \(O(N)\) action on the \(x^i\) where the radial constraint \(\mathbf{x}^2 = 1\) identifies the sphere with the coset space \(S^{N-1} = O(N)/O(N-1)\) (which is the Stiefel manifold \(V_1(\mathbb{R}^N)\)). Functions on the sphere are the induction of the trivial representation of \(O(N-1)\) into \(O(N)\), \(\text{Ind}_{O(N-1)}^{O(N)}\mathbf{1}\), whose decomposition into irreps of \(O(N)\) is
\begin{equation}\label{eq:d=2_stiefel_decomp}
  \text{Ind}_{O(N-1)}^{O(N)}\mathbf{1} = \bigoplus_{l=0}^{\infty}V_{(l,0,\dots,0)} .
\end{equation}
The irreps \(V_{(l,0,\dots,0)}\) are the ``spin \(l\)'' reps of \(O(N)\), \textit{i.e.} the traceless symmetric rank-\(l\) tensors, \textit{i.e.} the polynomials \(x_{\{i_1} \dots x_{i_l\}}\). These representations have eigenvalue \(l(l+N-2)\) under the \(O(N)\) quadratic Casimir. Note that these show up with unit multiplicity in the decomposition.

On the space of polynomials in the \(x_i\) there is a less obvious \(SL(2,\mathbb{R})\) action which commutes with the \(O(N)\). In particular, \(P = \mathbf{x}^2\), \(K = - \nabla^2\), and \(D =-i \big( x\cdot \pd + \frac{N}{2}\big)\) (which essentially measures the degree of homogeneous polynomials), close under commutation to give the \(\mathfrak{sl}(2,\mathbb{R})\) algebra. The \(SL(2,\mathbb{R})\) is the analog of the \(SU(2,2)\) in the main text, and we have named the above generators accordingly to highlight this fact. The harmonic polynomial condition---annihilation by \(K\)---determines the lowest weight state of an \(SL(2,\mathbb{R})\) representation, with the rest of the states in the representation obtained by repeated applicaiton of \(P\).

In complete analogy with the main text, specifying an \(O(N)\) harmonic polynomial specifies the \(SL(2,\mathbb{R})\) representation and vice-versa. This \(SL(2,\mathbb{R}) \times O(N)\) action is the prototypical example of a reductive dual pair~\cite{howe}.\footnote{The following appendix gives a more general introduction to reductive dual pairs.}

This \(SL(2,\mathbb{R}) \times O(N)\) duality is more than a simplified example of the \(SU(2,2) \times U(N)\) duality in \(d=4\) dimensions covered the main text. It is precisely the story for \(d = 2\) dimensions! To see this, recall that massless particles split into left- and right-movers in \(d = 1+1\), and that the (global) conformal group factorizes accordingly, \(SO(2,2) \simeq SL(2,\mathbb{R}) \times SL(2,\mathbb{R})\). Working with, say, the left movers, particles carry non-zero lightcone momentum \(p_i^- = p_i^0 - p_i^1\), where \(i = 1,\dots,N\) is a particle number index. In analogy to spinors, the ``square root of momentum'' in \(d = 1+1\) is just a simple change of variables, \(p_i^- = \la_i^2\).\footnote{We use \(\la\) for the variable name here just to emphasize the connection to the main text, but note that \(\la_i \in \mathbb{R}\) is not a spinor in the present context.} The construction of \(N\)-particle primaries for free theories in \(d=2\) thus boils down to finding the harmonic polynomials in the \(\la_i\), the solution to which we gave above.

%%%%%%%%%%%%%%%%%%%%%%%%%%%%%%
%%%%%%%%%%%%%%%%%%%%%%%%%%%%%%
%%%%%%%%%%%%%%%%%%%%%%%%%%%%%%
\section{The oscillator representation and conformal representations}
Here we provide some basic information about the oscillator representation, as well as some technical details about the conformal representations encountered in the main text. Among others, we have found the references~\cite{howe, howe1985, howe2012, kashiwara1978, Bargmann:1970, Woit:2017, Mack:1975, Luscher:1974, MackSalam, Ruehl:1973, Ruehl:1973pr} useful in helping us stitch together the following story.

Consider a set of \(n\) harmonic oscillators of unit mass and unit frequency. The Hamiltonian is
\begin{equation}
  H = \sum_{I=1}^n\frac{1}{2}\big(q_I^2 + \pi_I^2\big),
\end{equation}
with the coordinates obeying the usual canonical commutation relations (CCR)
\begin{equation}
  [q_I,\pi_J] = i\delta_{IJ}.
\end{equation}
Considered as a whole, the symplectic group \(Sp(2n,\mathbb{R})\) acts on phase space; it is an automorphism preserving the CCR,
\begin{equation}\label{eq:CCR_with_J}
  \Big[\begin{pmatrix} \mathbf{q} \\ \boldsymbol{\pi} \end{pmatrix}, \begin{pmatrix} \mathbf{q} \\ \bpi \end{pmatrix} \Big] = i J,~~~J = \begin{pmatrix} 0 & \mathbf{1} \\ -\mathbf{1} & 0 \end{pmatrix},
\end{equation}
where \(J\) is the usual symplectic bilinear form. As is readily checked, the generators of the symplectic algebra \(\mathfrak{sp}(2n,\mathbb{R})\) are the \textit{symmetric, quadratic} parts of the phase space coordinates,
\begin{equation}\label{eq:sp2n_gens}
 \frac{1}{2} \Big\{\begin{pmatrix} \mathbf{q} \\ \bpi \end{pmatrix}, \begin{pmatrix} \mathbf{q} \\ \bpi \end{pmatrix} \Big\} = \left \{ \def\arraystretch{1.2} \begin{array}{l} q_I q_J \\ \frac{1}{2} \{q_I,\pi_J\} \\ \pi_I\pi_J \end{array} \right. \,.
\end{equation}

As is standard material, the Hilbert space is spanned by the \(\ket{\mathbf{n}} = \ket{n_1\dots n_n}\) states. We can readily see this by passing to creation/annihilation operators,
\begin{equation}\label{eq:cayley}
  \begin{pmatrix} \mathbf{a} \\ \mathbf{a}^{\dag} \end{pmatrix} = \frac{1}{\sqrt{2}} \begin{pmatrix} \mathbf{1} & \mathbf{i} \\ \mathbf{1} & -\mathbf{i} \end{pmatrix} \begin{pmatrix} \mathbf{q} \\ \bpi \end{pmatrix},
\end{equation}
whereupon \([a_I,a^{\dag}_J] = \delta_{IJ}\) and \(H = \frac{1}{2} \sum_{I=1}^n \{a^{\dag}_I,a_I\} = \sum_{I=1}^n \big(a^{\dag}_I a_I + \frac{1}{2}\big) \). Analogous to eq.~\eqref{eq:sp2n_gens}, \(\frac{1}{2}\big\{ \big(\begin{smallmatrix} {\bf a} \\ {\bf a}^{\dag} \end{smallmatrix} \big),\big(\begin{smallmatrix} {\bf a} \\ {\bf a}^{\dag} \end{smallmatrix} \big) \big\}\) provide a (unitarily equivalent) realization of the symplectic algebra.

The Hamilitonian makes apparent that the energy eigenstates are degenerate and fall into representations of \(K= Sp(2n,\mathbb{R})\, \cap \, O(2n) \simeq U(n)\), which is the maximal compact subgroup of \(G=Sp(2n,\mathbb{R})\). The transformations in coset \(G/K\) mix states of different energies. In this way, the Hilbert space of the harmonic oscillators furnishes a representation of \(Sp(2n,\mathbb{R})\), known as the oscillator representation.\footnote{As a result of the \(1/2\)'s in the groud state energies, technically it may be a representation of the metaplectic group---the double cover of \(Sp(2n,\mathbb{R})\)---and hence is often called the metaplectic representation. This representation is also reducible, splitting into an even and odd component (schematically, this is because the generator \(a^{\dag}a^{\dag}\) causes jumps of two in the occupation number).}\,\footnote{Historically in physics, symmetries where the Hamiltonian was a generator in a larger symmetry group were called spectrum generating symmetries or dynamical symmetries.}

The abstract action of the symplectic group on the \(q\)s and \(\pi\)s (or \(a\)s and \(a^{\dag}\)s) is unitarily realized on a space of square integrable functions. The coordinate space wavefunctions \(\braket{\mathbf{q}|\mathbf{n}} = h_{\mathbf{n}}(\mathbf{q})e^{-\mathbf{q}^2/2}\), with \(h_{\mathbf{n}}(\mathbf{q}) = h_{n_1}(q_1)\cdots h_{n_n}(q_n)\) the appropriate product of Hermite polynomials, are a basis for this space with the standard inner product on \(\mathbb{R}^n\), \(\braket{\psi|\psi'} = \int d^nq \, \overline{\psi(q)}\psi'(q)\) with the bar denoting complex conjugation. Note that \(\pi_I \to -i \pd/\pd q_I\) is Hermitian with respect to this inner product. On the other hand, the creation/annihilation operators are realized on a complex space with basis \(\braket{\boldsymbol{\xi}|\mathbf{n}} = \boldsymbol{\xi}^{\mathbf{n}} = \xi_1^{n_1}\cdots \xi_n^{n_n}\) and inner product \(\braket{\psi|\psi'} = \int d^n\overline{\boldsymbol{\xi}} d^n\boldsymbol{\xi} \, e^{-\left|\boldsymbol{\xi}\right|^2} \overline{\psi(\boldsymbol{\xi})} \psi'(\boldsymbol{\xi})\). Note that with this measure \(a_I \to \pd/\pd \xi_I\) is the adjoint to \(a^{\dag}_I \to \xi_I\).

We call the representation in terms of \(q\)s and \(\pi\)s the Schr\"odinger representation, and that in terms of the \(a\)s and \(a^{\dag}\)s the Bargmann-Fock representation. The intertwining operator which maps us between these representations---\textit{i.e.} implements the transform in eq.~\eqref{eq:cayley}---is given by the Bargmann-Fock transform (in a more physics-oriented language, this is the operator \(\int d^n\overline{\boldsymbol{\xi}}d^n\boldsymbol{\xi} \braket{\mathbf{q}|\boldsymbol{\xi}} \bra{\boldsymbol{\xi}}\) and is commonly obtained via the introduction of coherent states). That such a map exists is guaranteed by the Stone-von Neumann theorem, which essentially states that there is only one unitary representation of the CCR, so that any two realizations must be unitarily equivalent.

The symplectic group contains many subgroups. A \textit{reductive dual pair} is a pair of groups \(G \times G' \subset Sp(2n,\mathbb{R})\) such that \(G\) and \(G'\) are maximal commutants with respect to one another, \textit{i.e.} \(G'\) is the largest subgroup in \(Sp(2n,\mathbb{R})\) that commutes with \(G\). When restricted to \(G \times G'\), the reduction of the oscillator representation of \(Sp(2n,\mathbb{R})\) is such that the irreps of \(G\) appearing determine those of \(G'\) and vice-versa. 

For our applications of decomposing the \(N\)-distinguishable-particle Hilbert space of free CFTs---\textit{i.e.} decomposing \(\mathcal{H}_N = \mathcal{H}_1^{\otimes N}\) into irreps of the conformal group---the relevant dual pairs are\footnote{It is possibly true that a similar mechanism works in \(d=6\).}
\begin{subequations}\label{eq:dualpairs}
  \begin{align}
    d=2:&~~ SL(2,\mathbb{R}) \times O(N) \subset Sp(2N,\mathbb{R}), \\
    d=3:&~~ Sp(4,\mathbb{R}) \times O(N) \subset Sp(4N,\mathbb{R}), \\
    d=4:&~~ SU(2,2) \times U(N) \subset Sp(8N,\mathbb{R}).
  \end{align}
\end{subequations}
In each case, the conformal primaries of \(\mathcal{H}_N\) are the harmonics of the relevant Stiefel manifold,
\begin{subequations}\label{eq:allstiefels}
  \begin{align}
    d=2:&~~ V_1(\mathbb{R}^N) \simeq O(N)/O(N-1), \\
    d=3:&~~ V_2(\mathbb{R}^N) \simeq O(N)/O(N-2), \\
    d=4:&~~ V_2(\mathbb{C}^N) \simeq U(N)/U(N-2).
  \end{align}
\end{subequations}
Analogous to eqs.~\eqref{eq:masterweight} and~\eqref{eq:d=2_stiefel_decomp}, for completeness we include the decomposition of functions on the Stiefel manifolds into irreps of \(O(N)\) or \(U(N)\), \textit{i.e.} the spectrum of primary operators in \(\mathcal{H}_N\):
\begin{subequations}\label{eq:allinduced}
  \begin{align}
    \text{Ind}_{O(N-1)}^{O(N)}\mathbf{1} &= \bigoplus_{l=0}^{\infty}V_{(l,0,\dots,0)}, \\
    \text{Ind}_{O(N-2)}^{O(N)}\mathbf{1} &= \bigoplus_{l_1\ge l_2} \bigoplus_{A=1}^{l_1-l_2+1} V^A_{(l_1,l_2,0,\dots,0)}, \\
    \text{Ind}_{U(N-2)}^{U(N)}\mathbf{1} &= \bigoplus_{l_1\ge l_2}\bigoplus_{\wtl_1\ge \wtl_2} %\bigg( \nonumber \\
    \bigoplus_{A=1}^{(l_1 - l_2 + 1)(\wtl_1 -\wtl_2 +1)}V^A_{(l_1,l_2,0,\dots,0,-\wtl_2,-\wtl_1)}. %\bigg).
  \end{align}
\end{subequations}
A few comments. As in the main text, we have labeled the finite-dimensional representations by their associated partitions/Young diagrams; recall that, in contrast to \(U(N)\), for \(O(N)\) these realize tensors which are traceless. For \(d=3,4\) the sum over \(A\) accounts for the multiplicity \(V_{L}\) shows up with in the decomposition; in accordance with the dual pair structure, it is simply the dimension of the Lorentz representation (\(SL(2,\mathbb{R})\) or \(SL(2,\mathbb{C})\) in \(d=3,4\), respectively). (The \(d=3\) spinorial realization of the conformal group is discussed below.) For \(d=3,4\), the above is strictly valid for \(N\ge 4\), with exceptional cases for \(N<4\)---see eq.~\eqref{eq:exceptional} for \(d=4\); the \(d=3\) analogue is straightforward to work out. For \(d=2,3\) these results follow from a decomposition using eq.~\eqref{eq:magic} like in eq.~\eqref{eq:ring_decomp} and the subsequent analysis. In \(d=4\), we had to identify the primary in a decomposition of a product of \(U(N)\) representations (see discussion following eq.~\eqref{eq:ring_decomp}); the analogue in \(d=2,3\) is identifying the primary in the restriction of a \(GL(N,\mathbb{R})\) representation labeled by partition \(L\) to \(O(N)\).\footnote{See, \textit{e.g.},~\cite{Bekaert:2006py} for a discussion of restricting \(GL(N)\) reps to \(O(N)\).}  Just as the \(d=4\) primary was the ``leading term'' in the decomposition, so it is in \(d=2,3\): the primary is the \(O(N)\) irrep labeled by the same partition \(L\) (but now, as an \(O(N)\) irrep, it corresponds to a traceless polynomial).
\begin{center}
\(\ast~~\ast~~\ast\)
\end{center}

With the above said, we now wish to make a few comments on the conformal representations encountered in the main text. We give this discussion in the context of the \(d=3\) case in order to provide some details for that story; the analogous statements for \(d=4\) (main text) and \(d=2\) (supplemental material on spherical harmonics) are easily generalized.

In \(d=3\) a massless momentum can be written as \(p_{ab} = \la_a\la_b\) with \(\la_a\) a two-dimensional real spinor. For \(N\) particles, the representation of the conformal algebra \(\mathfrak{sp}(4,\mathbb{R}) \simeq \mathfrak{so}(3,2)\) is given by
\begin{subequations}\label{eq:d=3_SR}
  \begin{align}
    P_{ab} &= \sum_i \la_a^i\la_b^i, \\
    K^{ab} &= -\sum_i \pd^a_i \pd^b_i, \\
    D &= \sum_i \frac{1}{2} \{\la_a^i,-i\pd^a_i\} \nonumber \\
    &= -i\frac{1}{2}\sum_i\big(\la^i\cdot \pd_i + 1\big), \\
    M_a^b &= \sum_i\frac{1}{2}\left(\{\la_a^i,-i\pd^b_i\} - \delta_a^b \frac{1}{2} \{\la_c^i,-i\pd^c_i\}\right) \nonumber \\
    &= -i\sum_i\left(\la_a^i\pd^b_i -\frac{1}{2}\delta_a^b \la^i\cdot \pd_i \right).
  \end{align}
\end{subequations}
This is the Schr\"odinger representation where the ``canonical momentum'' to the \(\la_a^i\) is \(\pi^a_i = -i \pd^a_i\).

For free QFT in \(d=3\) the scalar and fermion are the only massless particles which are conformal; respectively, they correspond to the trivial and non-trivial representation of the single particle little group \(O(N=1) \simeq Z_2\). As the theory is free, we can introduce fields containing both the positive- and negative-energy components, \textit{e.g.} for the scalar
\begin{equation}
  \phi(x) = \int \frac{d^2\la}{2(2\pi)^2} \left(e^{-\frac{i}{2}\la_a\la_bx^{ab}} A_{\la} + e^{\frac{i}{2}\la_a\la_bx^{ab}} A^{\dag}_{\la} \right),
\end{equation}
and for the fermion \(\psi_a(x) = \int \frac{d^2\la}{2(2\pi)^2}\big(\la_a e^{-\frac{i}{2}\la\la x}A_{\la} + \text{h.c.}\big)\).\footnote{The measure is simply the transform of \(\frac{d^3p}{(2\pi)^2}\delta(p^2){\theta(p^0)}\) into spinor variables; we note that the additional factor of \(1/2\) appearing from the Jacobian is \(1/\text{vol(little group)} = 1/\text{vol}(Z_2)\).} An operator is just some polynomial in fields and their derivatives; acting on the vacuum it creates a state
\begin{align}
  \mathcal{O}^{(N)}(x) \ket{0} = \int&\left( \prod_{i=1}^{N} \frac{d^2\la^i}{2(2\pi)^2} \right) e^{\frac{i}{2}\sum_i \la_a^i \la_b^i x^{ab}} \nonumber \\
  &\times f_{\scO}(\la_a^i) \ket{\la^1 \cdots \la^N}.
\end{align}
Here, we took \(\scO^{(N)}\) to contain \(N\) fields. \(f_{\scO}(\la)\) is some polynomial in the spinors; as discussed in the main text, if \(\scO^{(N)}\) is a conformal primary, then the polynomial is harmonic: \(K f_{\scO} = 0\). From here one can readily calculate 2-point functions (\(n\)-point functions follow with obvious generalization),
\begin{align}
  \langle \scO^{(N)}(x) \scO'^{(N')}(y) \rangle =& \delta_{N N'} \int \left( \prod_{i=1}^{N} \frac{d^2\la^i}{2(2\pi)^2} \right) \label{eq:SR_2pt} \\
  &\times e^{-\frac{i}{2}\sum_i\la_a^i \la_b^i (x-y)^{ab}} f_{\scO}(\la) f'_{\scO'}(\la). \nonumber
\end{align}

In order to discuss unitarity of the representation (and therefore the hermiticity of the generators in eq.~\eqref{eq:d=3_SR}), we need to define the inner-product. 
 A natural candidate is to define the in states at \(t = i\) where \(x^{ab} = i \delta^{ab}\) (and out states at the complex conjugate \(t = -i\)). At this point, the states have the appropriate Gaussian wave function factor for harmonic oscillators,
\begin{equation}\label{eq:quant_at_i}
  \braket{\la^1\cdots \la^N|\scO(t=i)} = f_{\scO}(\la) e^{-\frac{1}{2}\sum_{a,i}\la_a^i\la_a^i}.
\end{equation}
This is the scheme in~\cite{Luscher:1974, Mack:1975} and in modern literature this goes by the name ``NS quantization''~\cite{Rychkov:2016}; in addition to~\cite{Rychkov:2016} (the discussion of which is in Euclidean space), a good discussion in Lorentzian signature is the appendix of~\cite{Gillioz:2016}.

Imaginary time is familiar from the \(i\epsilon\) prescription to guarantee causality. However, it raises the question of where  the conformal representation actually ``lives''. Physically, we want quantum fields to live on Minkowski space, or its appropriate covering if we have non-integer scaling dimensions~\cite{Luscher:1974}. This is manifest in the standard construction of conformal representations as finite-component field representations \`a la Mack and Salam~\cite{MackSalam}, which constructs them as induced representations on \(G/\mathcal{P} = SO(d,2)/[ISO(d-1,1)\ltimes \mathbb{R}] \simeq \mathcal{M}^d\), where \(\mathcal{M}^d\) is \(d\)-dimensional Minkowski space and the parabolic subgroup is generated by Lorentz transformations, special conformal translations, and dilatations, which all preserve the origin of Minkowski space (\textit{i.e.} the coset is generated by translations, which obviously give all of Minkowski space).

On the other hand, it is natural to consider (induced) representations which live on \(G/K\), where \(K = SO(d)\times SO(2)\) is the maximal compact subgroup of the conformal group.\footnote{For the spinorial form of the conformal groups in eq.~\eqref{eq:dualpairs}, \(K = SO(2) \simeq U(1)\), \(Sp(4,\mathbb{R}) \cap O(4) \simeq U(2)\), and  \(S(U(2)\times U(2))\).} \(G/K\) is a Hermitian symmetric space; physically it is a certain analytic continuation of Minkowski space (\(\text{dim}(G/K) = 2d\)). In particular, it is an analogue of an ``upper-half space''. For example, in \(d=2\) \(SL(2,\mathbb{R})/SO(2) \simeq Sp(2,\mathbb{R})/SO(2) \simeq H^1\) \textit{is} the upper-half plane \(H^1\), while in \(d=3\) it is the Siegel upper-half space \(H^2\),
\begin{align}
  &\bigslant{Sp(4,\mathbb{R})}{Sp(4,\mathbb{R})\cap O(4)} \label{eq:siegel_H2} \\
  &~~~~~=\Big\{ z \in \text{Mat}_{2\times 2}(\mathbb{C}) ~:~ z = z^T,~ \text{Im}\,z > 0 \Big\} \,. \nonumber
\end{align}
That is, \(z^{ab} = x^{ab} + i y^{ab}\) consists of complex, symmetric \(2\times 2\) matrices with positive definite imaginary component.\footnote{For the spinorial conformal groups, \(G = SL(2,\mathbb{R})\), \(Sp(4,\mathbb{R})\), \(SU(2,2)\), \(G\) acts on (real or complexified) Minkowski space via fractional-linear transformations. That is, for \(g = \left( \begin{smallmatrix} A & B \\ C & D \end{smallmatrix} \right) \in G\) (with \(A..D\) \(2\times 2\) matrices for \(d=3,4\)), we have \(g: z \mapsto (Az + B)(Cz + D)^{-1}\). One readily verifies that this gives the familiar conformal transformations on the \(d\)-vector \(z^{\mu}\). {(Very schematically, Lorentz transformations and dilatations correspond to elements \( \left( \begin{smallmatrix} A &  \\  & A^{-1} \end{smallmatrix} \right)\), translations to \( \left( \begin{smallmatrix} 1 & B \\ 0 & 1 \end{smallmatrix} \right)\), and special conformal transformations to \( \left( \begin{smallmatrix} 1 & 0 \\ C & 1 \end{smallmatrix} \right)\); note that the groups may be generated from Lorentz transformations + dilatations + translations + the inversion element \(\left( \begin{smallmatrix} 0 & 1 \\ -1 & 0 \end{smallmatrix} \right)\) which takes \(z \to -z^{-1}\).)}} Importantly, Minkowski space lives at the boundary \(\text{Im}\,z \to 0\). These upper-half spaces are also known as ``tube domains''; they are the same tube domains appearing in \textit{e.g.}~\cite{streater}; here we see that they naturally arise as a homogeneous space of the conformal group.\footnote{For general \(d\), \(SO(d,2)/(SO(d) \times SO(2))\) basically complexifies Minkowski space via \(z^{\mu} = x^{\mu} + i y^{\mu}\) with \(y^{\mu}\) in the open forward light cone (\(y^0>0\), \(y^2 > 0\)).}

An important result, originally due to Harish-Chandra, is that when \(G/K\) is a Hermitian symmetric space, then \(G\) possesses holomorphic discrete series representations, which are \(K\)-valued functions that have an analytic expansion on all of \(G/K\). In particular, the function can be expanded in a Taylor series in \(z\), $\mathcal{O}(z) \sim \sum_k [\partial^k \mathcal{O}](0)z^k$. These are lowest-weight representations possessing an infinite tower of basis states obtained by applying the raising operators of the algebra. These lowest weight representations are analytic \textit{functions} on \(G/K\), and when we take them to the boundary they are realized as \textit{distributions} on Minkowski space, {\it e.g.}~\cite{Ruehl:1973,Ruehl:1973pr}.

Now, for the Schr\"odinger representation in eq.~\eqref{eq:d=3_SR} the generator
\begin{equation}
  \frac{\delta^{ab}}{2}\big(P + K\big)_{ab} = P^0 + K^0 = \sum_{i,a}\frac{1}{2}\big(\la_a^i\la_a^i + \pi^a_i\pi^a_i\big)
\end{equation}
is the Hamiltonian of the harmonic oscillators; the combination \(P^0+K^0\) is called the \textit{conformal Hamiltonian}~\cite{Luscher:1974, Mack:1975, Rychkov:2016}. Per our earlier discussion, it is obviously invariant under the maximal compact subgroup of the conformal group, \(K = Sp(4,\mathbb{R}) \cap O(4)\). The conformal group \(G = Sp(4,\mathbb{R})\) acts transitively on the upper half-space \(H^2\) in eq.~\eqref{eq:siegel_H2}; geometrically, \(K\) stabalizes the point \(i\left( \begin{smallmatrix} 1 & 0 \\ 0 & 1 \end{smallmatrix} \right)\) and the transformations in \(G/K\) act effectively, generating all of \(H^2\) from the point \(i \mathbf{1}\), whence eq.~\eqref{eq:quant_at_i}.

The upper half spaces under consideration are unbounded. They can be compactified by mapping them to generalized open unit disks. The map that implements this transformation is none other than the Cayley transform in eq.~\eqref{eq:cayley}. For example, in \(d=3\) we have
\begin{equation}
H^2 \to D^2 = \Big\{ z \in \text{Mat}_{2\times 2}(\mathbb{C}) ~:~ z = z^T,~ 1 - z^{\dag}z > 0\Big\}.
\end{equation}
Importantly, just as Minkowski space laid at the boundary of the upper-half space, at the boundary of the open disk sits conformally compactified Minkowski space.\footnote{{At the boundary \(z^\dag z = 1\), so \(z\) is a unitary, symmetric matrix. We can parameterize it as \(u \left( \begin{smallmatrix} t_1 + i t_2  &i t_3 \\ it_3 & t_1 - i t_2 \end{smallmatrix} \right)\) with \(\left| u\right| = 1\) a phase and \(t_i \in \mathbb{R}\) with \(t_1^2 + t_2^2 + t_3^2 = 1\). Hence, the boundary is topologically \(S^1 \times S^2\); for interacting theories the timelike circle needs to be unwrapped for obvious causality reasons, \(S^1 \to \mathbb{R}\)---in this way higher (possibly infinite) sheeted coverings of the conformal group come into play.}}

The passage to the Bargmann-Fock representation via the Cayley transform aligns closer to the typical way conformal representations are discussed in the modern literature with radial quantization in Euclidean space, \textit{e.g.}~\cite{Rychkov:2016,Simmons-Duffin:2016gjk}. In particular, the point \(i \mathbf{1}\) in the upper-half space is mapped to the origin of the unit disk, so the ``in-states'' lie at the origin. Moreover, the generators \(P_{\text{BF}} \sim a^{\dag}a^{\dag}\) and \(K_{\text{BF}} \sim a a\) are now adjoints of one another \(P_{\text{BF}}^{\dag} = K_{\text{BF}}\), while the dilatation operator \(D_{\text{BF}} \sim \sum \big(a^{\dag}a + \frac{1}{2} \big)\) is the Hamiltonian.\footnote{We emphasize that the special conformal generator is a generalized Laplacian in both the Schr\"odinger and the Bargmann-Fock representations, and so the harmonic polynomials take the exact same form in either representation.} 

One can, of course, also construct the inner product in the Bargmann-Fock representation and use this realization to compute \(n\)-point functions. This is straightforward. We wish to mention that in both the Schr\"odinger representation (eq.~\eqref{eq:SR_2pt} and its generalization to \(n\)-point functions) and the Bargmann-Fock representation, one is essentially dealing with Gaussian integrals. This is to be contrasted against the ``usual'' momentum variables---\textit{i.e.} no spinors---where we would be dealing with Fourier integrals \(e^{ip_{\mu}x^{\mu}}\). As physicists, Gaussian integrals are nicer as the machinery of Wick contractions and such is available. We also point out that the Bargmann-Fock representation may be even cleaner than the Schr\"odinger representation; although much remains to explore, at least for 2-point functions, many cross-terms in the expansion of the polynomials \(f_{\scO}f'_{\scO'}\) are trivially zero by homogeneity conditions (in the Schr\"odinger representation these cross-terms are generally non-zero, but the sum of all terms gives the appropriate cancellations).

As a final point, we discuss finite conformal transformations, \textit{i.e.} how the infinitesmal representation of the conformal algebra \(\mathfrak{g}\) integrates to a  representation of the group, \(G = \exp\big(i \mathfrak{g}\big)\). The finite transformations of translations, Lorentz transformations, and dilatations are straightforward. The tricky one is special conformal transformations.

Again, we proceed with our discussion in \(d=3\) where \(G = Sp(4,\mathbb{R})\). First, note that all of \(G\) may be generated from translations, Lorentz transformations, dilatations, and the element \(J = \left( \begin{smallmatrix} 0 & \mathbf{1} \\ -\mathbf{1} & 0 \end{smallmatrix} \right)\). \(J\) acts like the inversion element, sending Minkoski space coordinates \(J:~x \to -x^{-1}\). The action of \(J\) on the oscillator representation is easy to deduce. For \(N\) particles, \(J\) acts by (recall \(\pi^a_i = -i \pd^a_i\)) \(\left( \begin{smallmatrix} 0 & \mathbf{1} \\ -\mathbf{1} & 0 \end{smallmatrix} \right) \left( \begin{smallmatrix} \la^1 & \cdots & \la^N \\ \pi_1 & \cdots & \pi_N \end{smallmatrix} \right)\), so that \(J\) essentially exchanges \(\la\) and \(\pi\). In other words, up to a multiplicative constant, \(J\) is realized as a Fourier transform!\footnote{This is a unitary transformation by Plancherel's theorem. (In fact, logically speaking, we can use the oscillator representation to \textit{prove} unitarity of the Fourier transform~\cite{howe2012}.)} Recall that harmonic oscillator eigenstates---Hermite polynomials times a Gaussian factor---transform into themselves under Fourier transforms. An important fact, \textit{e.g.}~\cite{kashiwara1978}, is that for \(f(\la)e^{\frac{i}{2}\sum_i\la^i\la^ix}\) with \(f(\la)\) a harmonic polynomial, the Fourier transform returns a harmonic polynomial. This implies the correct transformation of a primary  operator \(\scO(x)\) in Minkowski space~\cite{kashiwara1978}.

To the best of our knowledge, it is not known what the unitary transformation/operator is which implements a finite special conformal transformation in momentum space for general (positive energy) representations of the conformal group.\footnote{The infintesmal transformations are well-known and easy to obtain: just Fourier transform the position space generators in~\cite{MackSalam}{, \textit{i.e.} \(-i\pd_x \to p\), \(x \to i \pd_p\)}. The infinitesmal form and their implied Ward identities has been the subject of intensive study in recent years, \textit{e.g.}~\cite{Coriano:2013, Bzowski:2013, Gillioz:2016}.} Above we explained how this transformation is realized on half-integer scaling dimension operators using spinors; we are hopeful that this may help in finding the finite transformations for general representations of the infinite-sheeted covering of the conformal group (\textit{i.e.} when there are anomalous dimensions, so the scaling dimension is not necessarily half-integer). In this spirit, we note that the connection of finite special conformal transformations---and, in fact, the whole setup of the oscillator representation---is reminiscent of twistors, where one might hope to interpret \(e^{\frac{i}{2}\la_ax^{ab}\la_b}\) as \(e^{i\la_a\pi^a}\) with \(\pi^a = x^{ab}\la_b/2\) giving some ``incidence relation''. Said another way, \(Sp(4,\mathbb{R})\) has an easy to understand linear action on \(\left( \begin{smallmatrix} \la \\ \pi \end{smallmatrix} \right)\), so that our setup feels quite close to a spinorial form of ``embedding space'' {\it e.g.}~\cite{Rychkov:2016} (a spinorial form of the projective null cone), and twistors are probably the natural language to formalize this.
 
%%%%%%%%%%%%%%%%%%%%%%%%
%%%%%%%%%%%%%%%%%%%%%%%%
%%%%%%%%%%%%%%%%%%%%%%%%

\section{Spin current}
Here we elaborate on \textit{(i)} a convenient form for manipulating the currents \(J^{(n,m)}_{1^{n+m-k}2^k}\) and \textit{(ii)} a mathematical reason---Frobenius reciprocity---for why the currents must be conserved.

As all of the Lorentz indices are symmetrized, to simplify matters we introduce an index-free notation wherein we contract the \(J^{(n,m)}\), \(f^{(n,m)}\), \textit{etc}. with auxiliary spinors \(\xi^a\) and \(\xit^{\da}\) (similar constructs are frequently used in the CFT bootstrap literature, \textit{e.g.}~\cite{Costa:2011mg}). We denote contracted objects with a hat, \textit{e.g.}
\begin{equation}
  \widehat{f}^{(n,m)} \equiv \xi^{a_1}\cdots \xi^{a_n}\xit^{\da_1} \cdots \xit^{\da_m} f^{(n,m)}_{a_1\dots a_n\da_1 \dots \da_m} .
\end{equation}
The uncontracted object is recovered simply by differentiating,
\begin{equation}
  f^{(n,m)}_{a_1 \dots a_n \da_1 \dots \da_m} = \frac{1}{n!} \left(\frac{\pd}{\pd \xi^a}\right)^n \frac{1}{m!}\left(\frac{\pd}{\pd \xit^{\da}}\right)^m \widehat{f}^{(n,m)},
\end{equation}
which ultimately is the same as instructing one to symmetrize the object into which the auxiliarly spinors are contracted, 
\begin{equation}
  \frac{1}{n!} \left(\frac{\pd}{\pd \xi^a}\right)^n \big(\xi^b\big)^n = \delta_{(a_1}^{b_1} \cdots \delta_{a_n)}^{b_n}.
\end{equation}
For a simpler notation, we further define
\begin{align}
  x &\equiv \widehat{\la} \equiv \xi^a \la_a,~~~ y\equiv \widehat{\et} \equiv \xi^a \et_a, \nonumber \\
  \overline{x} &\equiv \widehat{\lt} \equiv \xit^{\da} \lt_{\da},~~~ \overline{y}\equiv \widehat{\etl} \equiv \xit^{\da} \etl_{\da}.
\end{align}

With these preliminaries, \(\widehat{f}^{(n,m)}\) takes the form of a simple polynomial
\begin{subequations}\label{eq:fhat_expansion}
\begin{align}
  \widehat{f}^{(n,m)} &= (c x + y)^n (\overline{y} - \tfrac{1}{c}\overline{x})^n \\
  &= \sum_{k=0}^{n+m} c^{n-k} \binom{n+m}{k}\widehat{J}^{(n,m)}_{1^{n+m-k}2^k},
\end{align}
\end{subequations}
where the second equality comes from eq.~\eqref{eq:fnm_expansion}. By using the binomial expansion \((a+b)^n = \sum_{i=0}\binom{n}{i} a^{n-i}b^i\), one can expand \(\widehat{f}^{(n,m)}\) and match powers of \(c^{\#}\) to obtain an expression for the currents \(\widehat{J}^{(n,m)}_{1^{n+m-k}2^k}\). The result is a piecewise expression, which can be written as a hypergeometric series. This is not particularly illuminating, so we omit the formula. More to the point, a main purpose of this discussion was to introduce the index-free notation, which often renders expressions more amenable for computation.

Now we proceed to give a mathematical justification for why the non-holomorphic primaries in \(\scH_2\) must be conserved currents. This is a consequence of Frobenius reciprocity, which roughly codifies the idea that induction and restriction are inverse actions. In more detail, consider the induced representation \(\text{Ind}_H^GW\) of a representation \(W\) of \(H \subset G\) (\textit{i.e.} a \(H\)-valued function on the coset \(G/H\)). For compact groups, the finite-dimensional irreps of \(G\) provide a basis for such functions. Frobenius reciprocity states that the multiplicity of a representation \(V\) of \(G\) appearing in \(\text{Ind}_H^GW\) is equal to the multiplicity for which \(W\) appears in the restriction of \(V\) to \(H\): \(\text{mult}(V,\text{Ind}_H^GW) = \text{mult}(W,\text{Res}_H^GV)\).\footnote{A simple example are spherical harmonics. The traceless symmetric tensors \((l,0,\dots,0)\) of \(O(N)\) all contain the trivial representation of \(O(N-1)\) exactly once, \textit{e.g.} an \(O(N)\) vector splits into a vector plus a singlet of \(O(N-1)\). This reasoning is reflected in the multiplicity free decomposition in eq.~\eqref{eq:d=2_stiefel_decomp}.} For the problems we consider, this multiplicity is reflected that the harmonic polynomials carry both \(U(N)\) indices and Lorentz indices, \textit{i.e.} the multiplicity is equal to the dimension of the Lorentz representation.

For the currents in the main text, these are in \(U(2)\) representations labeled by Young diagrams with \(n+m\) boxes. Moreover, the Stiefel manifold is just \(U(2)\) in this case, \(U(N)/U(N-2) \to U(2)/1 = U(2)\). Therefore, since the restriction of any representation to the trivial group \(H = 1\) is simply the dimension of the representation, Frobenius reciprocity tells us that the multiplicity that this representation appears with is \(n+m+1\). At first glance it seems there is a paradox since the dimension of the \(SL(2,\mathbb{C})\) representation is \((n+1)(m+1)\). This is resolved by the fact that the currents are conserved: \(\big(P\cdot J^{(n,m)}_{1^{n+m-k}2^k}\big)_{a_1\dots a_{n-1} \da_1 \dots \da_{m-1}}\) transforms in the spin \((\frac{n-1}{2},\frac{m-1}{2})\) rep of \(SL(2,\mathbb{C})\). Hene \(P \cdot  J^{(n,m)}_{1^{n+m-k}2^k} = 0\) gives \(\text{dim}(\frac{n-1}{2},\frac{m-1}{2}) = nm\) constraints, leaving \((n+1)(m+1) - nm = n+m+1\) independent components.

\end{document}